%% file: ad2_mnras.tex
\documentclass[fleqn,usenatbib]{mnras}
\usepackage{newtxtext,newtxmath}
\usepackage[T1]{fontenc}
\usepackage{lipsum}
\DeclareRobustCommand{\VAN}[3]{#2}
\let\VANthebibliography\thebibliography
\def\thebibliography{\DeclareRobustCommand{\VAN}[3]{##3}\VANthebibliography}

\usepackage{amsmath,xspace}
\usepackage{graphicx}
\usepackage{lineno}

\DeclareMathOperator{\sech}{sech}
\DeclareMathOperator{\csch}{csch}
\newcommand{\Msol}{\ensuremath{\mathrm{M}_{\odot}}\xspace}
\newcommand{\kms}{\ensuremath{\mathrm{km\ s^{-1}}}}
\newcommand{\pPXF}{\textsc pPXF ~}
\newcommand{\dnk}{\ensuremath{\mathrm{D_n(4000)}}}
\newcommand{\tlw}{\ensuremath{t_\mathrm{LW}}}

\title[Asymmetric drift in MaNGA disks]{Asymmetric drift in MaNGA:
  Mass and radially-dependent stratification rates in galaxy disks}

\author[M. A. Bershady et al.]{
Matthew A. Bershady,$^{1}$\thanks{E-mail: mab@astro.wisc.edu},
Kyle B. Westfall,$^2$ 
Shravan Shetty,$^{1,3}$, 
David R. Law$^4$, 
Michele Cappellari$^5$,
\newauthor
Niv Drory$^6$, 
Kevin Bundy$^{2,7}$, 
Renbin Yan$^{8}$
\\
$^1$University of Wisconsin, Department of Astronomy, 475 N. Charter St., Madison, WI 53706, USA\\
$^2$University of California Observatories, University of California,
Santa Cruz, 1156 High St., Santa Cruz, CA 95064, USA\\
$^3$Kavli Institute for Astronomy and Astrophysics, Peking University, Beijing 100871, China\\
$^4$Space Telescope Science Institute, 3700 San Martin Drive, Bal- timore, MD 21218, USA\\
$^5$Sub-department of Astrophysics, Department of Physics, Uni- versity of Oxford, Denys Wilkinson Building, Keble Road, Oxford OX1 3RH, UK\\
$^6$McDonald Observatory, The University of Texas at Austin, 2515 Speedway, Stop C1402, Austin, TX 78712, USA\\
$^7$Department of Astronomy and Astrophysics, University of California, Santa Cruz, 1156 High St., Santa Cruz, CA 95064, USA\\
$^{8}$Department of Physics, The Chinese University of Hong Kong, Shatin, N.T., Hong Kong S.A.R., China
}

\date{Accepted XXX. Received YYY; in original form ZZZ}

\pubyear{2023}

\begin{document}
\label{firstpage}
\pagerange{\pageref{firstpage}--\pageref{lastpage}}
\maketitle

\begin{abstract} 

We measure the age-velocity relationship from the lag between ionized gas and stellar tangential speeds in  $\sim$500 nearby disk galaxies from MaNGA in SDSS-IV. Selected galaxies are kinematically axisymmetric. Velocity lags are asymmetric drift, seen in the Milky Way's (MW) solar neighborhood and other Local Group galaxies; their amplitude correlates with stellar population age. The trend is qualitatively consistent in \textit{rate} ($\dot{\sigma}$) with a simple power-law model where $\sigma~\propto~t^b$ that explains the dynamical phase-space stratification in the solar neighborhood. The model is generalized based on disk dynamical times to other radii and other galaxies. We find in-plane radial stratification parameters $\sigma_{0,r}$ (dispersion of the youngest populations) in the range of 10-40 \kms~ and $0.2<b_r<0.5$ for MaNGA galaxies. Overall $b_r$ \textit{increases} with galaxy mass,  \textit{decreases} with radius for galaxies above 10.4 dex (M$_\odot$) in stellar mass, but is $\sim$constant with radius at lower mass. The measurement scatter indicates the stratification model is too simple to capture the complexity seen in the data, unsurprising given the many possible astrophysical processes that may lead to stellar population dynamical stratification. Nonetheless, the data show dynamical stratification is broadly present in the galaxy population, with systematic trends in mass and density. The amplitude of the asymmetric drift signal is larger for the MaNGA sample than the MW, and better represented in the mean by what is observed in the disks of M31 and M33. Either typical disks have higher surface-density or, more likely, are dynamically hotter (hence thicker) than the MW. 

\end{abstract}



\input{introduction}

\input{data_sample}

\input{measurements}

\input{analysis}

\input{local_group}

\input{conclusion}

\section*{Acknowledgements}
    
This research was directly supported by the
  U.S. National Science Foundation (NSF) AST-1517006 and South African National Research Foundation (NRF) SARCI/114555.

Funding for the Sloan Digital Sky Survey IV has been provided by the
Alfred P. Sloan Foundation, the U.S. Department of Energy Office of
Science, and the Participating Institutions. SDSS-IV acknowledges
support and resources from the Center for High-Performance Computing
at the University of Utah. The SDSS web site is www.sdss.org.

SDSS-IV is managed by the Astrophysical Research Consortium for the
Participating Institutions of the SDSS Collaboration including the
Brazilian Participation Group, the Carnegie Institution for Science,
Carnegie Mellon University, the Chilean Participation Group, the
French Participation Group, Harvard-Smithsonian Center for
Astrophysics, Instituto de Astrof\'isica de Canarias, The Johns
Hopkins University, Kavli Institute for the Physics and Mathematics of
the Universe (IPMU) / University of Tokyo, Lawrence Berkeley National
Laboratory, Leibniz Institut f\"ur Astrophysik Potsdam (AIP),
Max-Planck-Institut f\"ur Astronomie (MPIA Heidelberg),
Max-Planck-Institut f\"ur Astrophysik (MPA Garching),
Max-Planck-Institut f\"ur Extraterrestrische Physik (MPE), National
Astronomical Observatories of China, New Mexico State University, New
York University, University of Notre Dame, Observat\'ario Nacional /
MCTI, The Ohio State University, Pennsylvania State University,
Shanghai Astronomical Observatory, United Kingdom Participation Group,
Universidad Nacional Aut\'onoma de M\'exico, University of Arizona,
University of Colorado Boulder, University of Oxford, University of
Portsmouth, University of Utah, University of Virginia, University of
Washington, University of Wisconsin, Vanderbilt University, and Yale
University.

\section*{Data Availability}

The data underlying this article are available in the Sloan Digital Sky Survey IV Data Release 17 (https://www.sdss4.org/dr17/).

\bibliographystyle{mnras}
\bibliography{ad}

\appendix

\input{appendix_apertures}

\input{appendix_age}

\input{appendix_ad_svde}

\input{appendix_lwa_systematics}

\bsp 
\label{lastpage}
\end{document}

%% file: introduction.tex
\section{Introduction}
\label{sec:introduction}

The dynamical structure of stellar populations in the MW disk is characterized by the age-velocity-dispersion relation (AVR), whereby older disk stars are dynamically hotter than their younger cohort, and hence have greater vertical extent. Knowledge of this relationship stems back nearly a century to observations by \cite{Stromberg1925} of the varying scale-heights of stars of different types in the solar neighborhood \citep[and substantially updated and refined by e.g.,][]{Edvardsson93, Dehnen98, Nordstrom04, Holmberg07,Sharma21}. 

AVR is a dynamical stratification of stars in the phase space of position and velocity, and as such AVR has direct relevance to understanding how stellar disks form and evolve \citep{Seabroke07,Aumer09}. The processes leading to this stratification have been debated for over 70 years. They include dynamical heating from two-body scattering of stars in dynamically cold stellar disks with in-plane resonances (spiral arms and bars) or giant molecular clouds \citep{Spitzer51, Spitzer53, Kokubo92, Carlberg85}; stirring by compact halo objects, satellites or mergers \citep{Toth92, Walker96, Huang97, Abadi03, Benson04, House11, Helmi12, Few12, RuizLara16, Pinna19}; or a relic of a dynamically hotter distribution of bulk velocities of star-forming clouds at earlier epochs \citep[][commonlly referred to as gas cooling]{Brook04, Bournaud09, Forbes12}. Recent models often include both these `heating' and `cooling' scenarios in the formation of disk dynamical stratification \citep[e.g.,][]{Martig14b, Navarro18, Bird21, McCluskey23}. Whether the AVR represents gas cooling, dynamical heating, or both, very basic questions remain as to whether AVRs exist broadly in the external galaxy population, and if so, how do they compare to what we observe in the MW?

Measuring AVR in resolved stellar populations currently is limited to galaxies in the Local Group \citep[LG][]{Collins2011, Dorman15, Beasley15, Bhattacharya2019, Quirk19, Quirk22}. Based on a compilation of eight LG galaxies, only two of which have stellar masses above 10$^{10}$ M$_\odot$, \cite{Leaman17} have suggested there is an interplay between heating mechanisms that correlates with galaxy mass. It is also abundantly clear from multi-color images of edge-on galaxies that star-formation and young stellar populations are confined mostly to disk mid-planes at least in more massive spirals, while older stellar populations are evident in thicker disks that extend beyond the central dust layer. In lower mass disks (galaxies with $V_c<120$ \kms, or below about $\log {\rm M}_\star / {\rm M}_\odot < 10.3\pm0.1$ dex in stellar mass, scaling ${\rm M}_\star$ from the MW), however, the situation appears to be different at least insofar as dust lanes (and therefore one may presume star-forming regions) are not as thinly distributed \citep{Dalcanton04}. Progress is needed beyond these qualitative statements and AVR measurements solely in the LG.

In principle AVR can be measured directly in integrated starlight with ample spectral resolution to measure integrated absorption line-widths in dynamically cold systems, and ample signal-to-noise (SNR) to disentangle different population components. Unfortunately, the necessary telescopes and instrumentation for such measurements is limited, and hence so too are the available data. However, it is possible to use velocities alone to estimate stellar dispersions by employing the well-known lag between the tangential speed of stars and the circular speed of the potential. This lag is known as asymmetric drift (AD). In integrated star-light, the challenge has still been to determine the stellar kinematics of multiple age components. \cite{Shetty20} pioneered such a method in a small sample of galaxies with optical integral-field spectroscopy (IFS). The results show great promise but the method is computationally expensive and requires exquisite SNR, as do other methods employing higher moments of the velocity field \citep{Coccato18,Poci19}.

In this paper we step back to measure AD in single (light-weighted mean) stellar components. We take advantage of very large IFS surveys to leverage the dynamic range in mean stellar population ages between different galaxies and different regions within galaxies. We infer AVR from stellar population age and AD measurements in the context of a simple, time-continuous stratification model. In so doing we recognize two basic assumptions inherent to this approach, namely (i) that the stratification process between galaxies and different regions of individual galaxies are the same; and (ii) that the stratification rate is continuous in time. We expect these assumptions are at best approximations, so we explore radial and mass dependencies of AVR model parameters to test them.

This paper uses optical IFS of nearby galaxies from the MaNGA survey \citep{Bundy15,Yan16b} in SDSS-IV \citep{Blanton17}. MaNGA uses fiber integral-field units \citep{Drory15} with the BOSS spectrographs \citep{Smee13} on the Sloan 2.5m telescope \citep{Gunn06}. The MaNGA observing strategy, target selection, spectrophotometric calibration and data reduction pipeline are described in \citet{Law15}, \citet{Wake17}, \citet{Yan16a}, and \citet{Law16_DRP}, respectively. Salient features of the spectroscopic data are discussed in relevant sections below. We take advantage of existing kinematic measurements of stars and gas from the Data Analysis Pipeline \citep[DAP;][]{Westfall19,Belfiore19} to select targets and define kinematic geometries.

The paper is structured as follows. The MaNGA data and sample selection is presented in Section~\ref{sec:data_sample}. Measurements that define the kinematics to compute asymmetric drift and spectral indices to estimate stellar populations ages in integrated starlight are described in Section~\ref{sec:measure} and Appendix~\ref{app:app}. The definition for asymmetric drift is found in Section~\ref{sec:AD_SVDE}. A correction to the gas tangential speed, due to its finite pressure support, to estimate the circular speed of the potential is detailed in Section~\ref{sec:gas_cor}; notably, this method does not rely on direct measurements of the gas dispersion in low signal-to-noise data well below the instrumental spectral resolution. A robust and reproducible method to determine stellar population ages based on \dnk~ is defined in Section~\ref{sec:ages}, with further details provided in Appendix~\ref{app:age}. 

\begin{figure*}
  \centering
  \includegraphics[width=\linewidth]{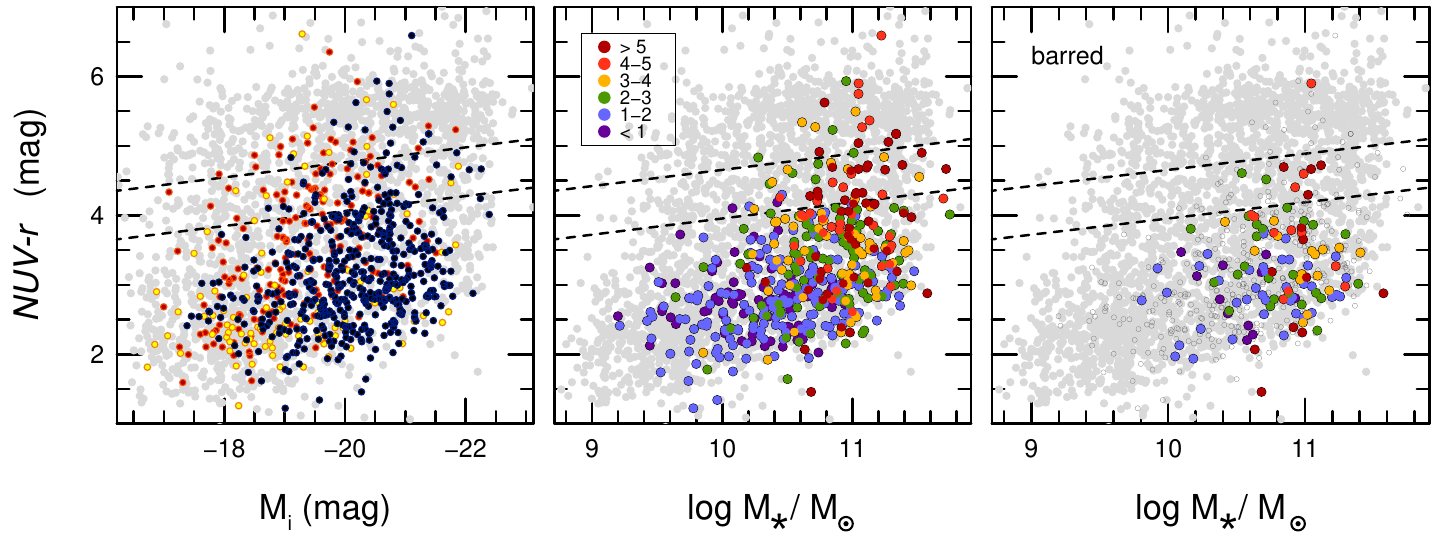}
  \caption{Rest-frame $NUV-r$ versus $M_i$ (left) and stellar mass (middle and right). The left panel shows the MPL-5 MaNGA sample of 2744 galaxies (grey points). The 798 kinematically regular galaxies (selected as described in text) are marked for the final subset of 497 galaxies (dark points), 192 galaxies excluded for being too inclined (red points), and 109 galaxies excluded for having too low S/N in either the stellar or ionized gas velocity fields (yellow points). In the middle panel the 497 galaxies are color coded by Sersic index as given in the key. Galaxies with strong bar morphology are plotted with the same color-coding in the right panel. Two parallel dashed lines delineate the blue cloud, green valley and red sequence.}
  \label{fig:cmd}
\end{figure*}

A simple power-law stratification model is motivated in Section \ref{sec:stratification}, starting with the relation of AD to in-plane components of the stellar velocity dispersion ellipsoid (SVE; \ref{sec:sAD_to_sr}), with supporting derivations found in Appendix~\ref{app:CBE}. The parameters of this stratification model are the radial components of the initial velocity dispersion at zero age ($\sigma_{0,r}$) and the power-law stratification exponent $b_r$.  In this context, analysis of our age and asymmetric drift measurements are presented in Section~\ref{sec:analysis}. Corrections for the different light-weighting for stellar population ages and kinematics have been extensively modeled, with modest corrections estimated and applied, as summarized in Section~\ref{sec:lwa_sys} and detailed in Appendix~\ref{app:lwa_sys}. Application of the stratification model to the data, and resulting values for the stratification parameters as a function of radius and mass can be found in Sections~\ref{sec:rad_trend} and \ref{sec:strat_var} respectively. We examine the impact of bars on the derived stratification parameters in Sections~\ref{sec:rad_trend} as well. For unbarred galaxies, the covariance between the stratification parameters and correlation with other physical parameters, such as radial light concentration, are found in Section~\ref{sec:discussion}. Finally, in Section~\ref{sec:LG} we make a comparison of the stratification parameters derived for MaNGA galaxies with those for the three most massive galaxies in the Local Group: M31, MW, M33. Our methods and findings are consolidated in a summary, Section~\ref{sec:conclusion}. Where applicable we adopt a flat $\Lambda$CDM cosmology with $\Omega_M = 0.3$ and $H_0 = 70~\kms~{\rm Mpc}^{-1}$. All magnitudes are in the AB system.

%% file: data_sample.tex
\section{Data Set and Sample Selection}
\label{sec:data_sample}

\subsection{Data}

The kinematic and spectrophotometric measurements used in this paper are based on the final SDSS-IV Data Release 17 \citep[DR17;][]{DR17} products for MaNGA described in more detail in papers describing the DAP \cite{Westfall19,Belfiore19}. 
The spectral resolution is $\sim$67~\kms\ ($\sigma$). The median spatial resolution is $1.4$ arcsec, varying as a function of mass given the MaNGA sample selection \citep{Wake17}; 75\% of the sample has a resolution between 1.1 and 2.1 arcsec, while 90\% of the sample has a resolution between 0.8 and 3.4 arcsec. Most relevant for the current analysis is that the distribution of apparent half-light radii is independence of mass. Given the extent of the analysis involved, however, the sample is limited to the set of galaxies processed in SDSS-IV Data Release 14 \citep{DR14}, which consists of 2744 unique observations of galaxies with MaNGA IFUs. Because we are interested in measuring kinematics in specific geometric regions of each galaxy (see Section~\ref{sec:measure}) we eschew the Voronoi-binned data and use only the spaxel-based DAP data products, which for DR17 correspond to the data-type SPX-MILESHC-MASTARSSP. For this data-type the stellar and ionized gas kinematics were determined with \pPXF \citep{Cappellari2017} and a hierarchically-clustered \citep[Sec.~5]{Westfall19} set of template stars from the Miles \citep{Falcon-Barroso11} stellar library.\footnote{This results in 42 templates, with no preselection on age or metallicity.} One of the advantages of relying only on velocities (versus higher moments) for our analysis is that the details of the stellar template constructions are relatively unimportant. We also use the DAP's measure of \dnk~ \citep[as defined by][]{Balogh99} and ionized gas fluxes. In addition to the DAP, we made use of the photometry (we adopt Petrosian magnitudes) and Sersic index provided by the NASA-Sloan Atlas \citep[NSA,][]{Blanton11}. 

\subsection{Sample Selection}

\begin{figure*}
  \centering
  \includegraphics[width=1\linewidth]{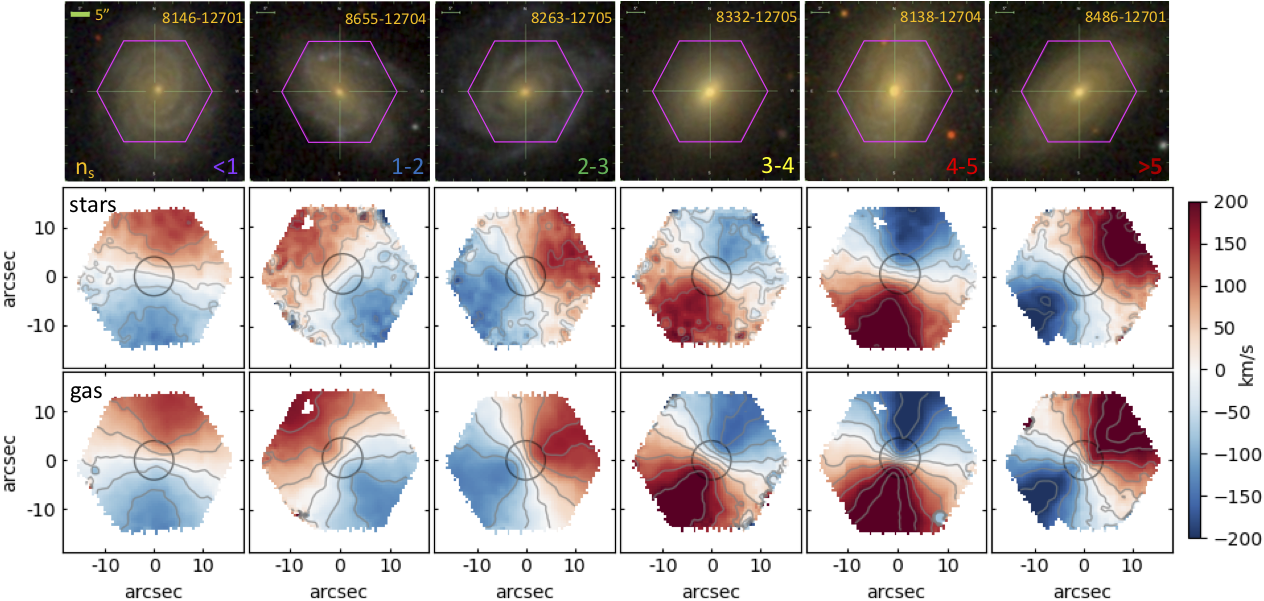}
  \caption{Examples of sample galaxies as a function of increasing Sersic index (labeled, left to right) \textit{without} strong bar morphology. Purple hexagons indicate the MaNGA IFU footprint, with a green horizontal bar in the top-left panel showing the angular scale of 5 arcsec. Galaxy IFU and plate numbers are marked. Stellar and gas velocity fields are shown in the middle and bottom rows. Dark-grey open circles indicate a radius of 4 arcsec, within which we do not consider kinematics measurements due to the strong impact of beam-smearing on the line-of-sight measurements.}
  \label{fig:examples_nobar}
\end{figure*}

We expect AD to be measurable in all galaxies with detectable dense gas,
but if significant kinematic irregularities or asymmetries are
present, this likely prevents the measurement of an accurate AD
signal. The DAP kinematics were used to select galaxies on the basis
of the regularity of both stellar and ionized gas velocity fields. To
make this assessment for such a large galaxy sample, and in order to
be quantitative, we fit simple kinematic models to all 2744 galaxies
in the DR14 MaNGA sample in an automated way.  The kinematic fitting
is based on the scheme developed by \cite{Westfall11} and
\cite{Andersen2013} using a single, inclined disk model with smooth
axisymmetric tangential motion and a radial trend described by $V(r) =
V_{\rm rot}~\tanh( R/h_{\rm rot} )$. The fitting solves for the
geometric and kinematic parameters.  The scheme is extended here to
fit ionized gas and stellar velocity fields simultaneously with the same
geometry (center, inclination, position angle) while allowing for the
rotation curve parameters ($V_{\rm rot}, h_{\rm rot}$) and systemic
velocity to be independent. We perform 12 velocity-field fits for each
galaxy: (1) four fits are permutations of fixing and freeing the
center and inclination and (2) each of these four fits is done when
only fitting the ionized gas data, only fitting the stellar data, and
fitting both tracers simultaneously. Fits with fixed geometric
parameters adopt photometric values parameters provided by the NSA.

The purpose of the multiple fitting permutations is to assess the
stability of both the kinematic and geometric parameters constrained
by different sets of tracers. Based on visual inspection of a subset
of cases for a range of variation in fitted parameters we assessed
quantitative limits to derive a kinematically regular sample.  We find
it is useful to limit variation between permutations and tracers on
the systemic velocity, kinematic center, position angle, and
inclination. Our AD-$\sigma$ analysis is of galaxies that satisfy the
following set of constraints on the results from the velocity field
fitting:

(i) When the two velocity fields are fitted simultaneously, the differences between the systemic velocities fit to the ionized gas and stellar velocity fields must be within $\sim7$ \kms~ (absolute value) for all four of the fit permutations.

(ii) In the two simultaneous fits to the ionized gas and stellar data
with the dynamical center left free, the dynamical center must be
within half of the imposed bounding box on the morphological center.
The bounding box is set to $\pm$(1,2,2,3,3) arcsec for
IFUs with (19,37,61,91,127) fibers respectively.

(iii) For all the relevant fit permutations, the difference between
the ionied-gas-only and stellar-only fitted position angles must be
less than 15$^\circ$.

(iv) In all six fits when the inclination is freely fit, the
best-fitting kinematic inclination must be in the range $15^\circ <
i_{\rm kin} < 80^\circ$. The difference between the ionized-gas-only and
stellar only kinematic inclinations must be less than $20^\circ$ both
with and without a fixed center. When both tracers are fit
simultaneously, the kinematic inclination must also be within
$20^\circ$ of photometric inclination derived from the standard
formula using $\epsilon = 1 - b/a$ and an intrinsic oblateness $q_0 =
1/8$ characteristic of spiral galaxies \citep[e.g.,][]{Padilla2008}: $\cos^2 i_{\rm phot} = [(1-\epsilon^2)-q_0^2]/(1-q_0^2)$.

\begin{figure*}
  \centering
    \includegraphics[width=1\linewidth]{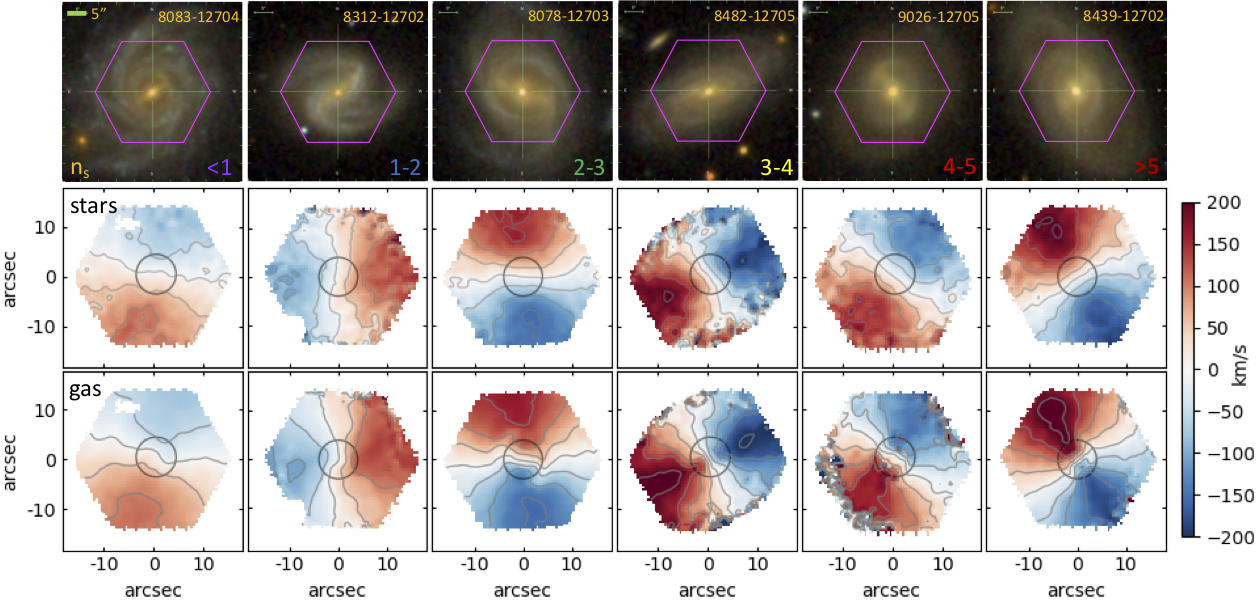}
  \caption{Examples of sample galaxies with bars as a function of increasing
    Sersic index, formatted as in Figure~\ref{fig:examples_nobar}.}
  \label{fig:examples_bar}
\end{figure*}

The simultaneous application of all these cuts yields a sample of 798 galaxies with regular kinematics. Visual inspection of these  galaxies resulted in a further culling to eliminate 192 objects that appear too inclined for reliable kinematics and colors due to uncertainties in internal extinction corrections. This visual inspection removes a subset of galaxies with a median inclination of $\sim66^\circ$, including all galaxies with photometric inclinations above 70$^\circ$. The same visual inspection also identified an additional 109 objects for removal because they had either too low in continuum or line S/N to have adequately sampled stellar or ionized gas velocity fields, respectively, despite the fitting process described above. The following analysis uses the remaining 497 galaxies (18\% of the parent sample).

The distribution of this sample in relation to the parent sample in the $NUV-r$ -- $M_i$ color-magnitude diagram (CMD) is shown in Figure~\ref{fig:cmd}.  Unsurprisingly, highly-inclined systems are preferentially at redder colors and lower absolute magnitude due to uncorrected internal-extinction. The low-S/N galaxies tend to be found at the peripheries of the CMD distribution at low luminosity or red color. Examples of the galaxy morphology are shown in Figure~\ref{fig:examples_nobar} and \ref{fig:examples_bar} for a range of Sersic index, unbarred and barred (based on our visual inspection). These Figures shows that in our sample the high-index galaxies still have disks, concluding that the high-index values are due to the light-weighted nature of the one-zone Sersic model, i.e., the high Sersic indices are a measure of the profile shape of the inner light concentration. 

Figure~\ref{fig:cmd} also shows that barred and unbarred galaxies in our sample have similar distributions in color total stellar mass. Despite the kinematics of the barred galaxies having sufficient regularity to meet our sample selection criteria, the bar perturbations to the velocity field are evident, particularly at small radii in Figure~\ref{fig:examples_bar}. Since our measurements avoid the inner regions due to beam smearing, described in the following section, a priori it was not clear if the bar perturbations would be problematic to our analysis. For example, \cite{Stark18} found inner position-angle twists that appeared to be associated with bars typically occurred within $r_{50}/2$, where $r_{50}$ is the half-light radius; this radius is within the smallest radial bin we consider (Section~\ref{sec:measure}). We therefore initially retain the barred sample with the intent to determine if there are systematic differences in the stratification properties in barred and unbarred galaxies. In the end, our expectations turn out to be too optimistic; we eliminate barred galaxies (roughly a quarter of the sample) in our tertiary analysis (Section~\ref{sec:ensemble_vs_indy} and onward).

%% file: measurements.tex
\section{Kinematic and spectrophotometric measurements}
\label{sec:measure}

\begin{figure}
  \centering
  \includegraphics[width=1.0\columnwidth]{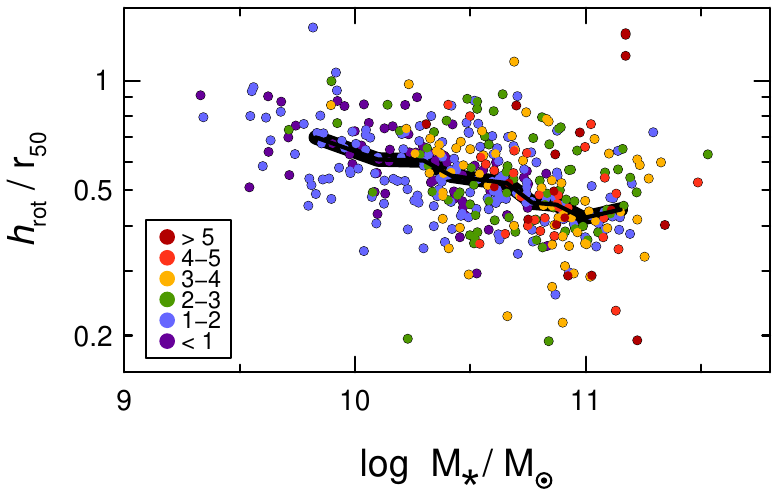} 
  \caption{Distribution of the rotation-curve rise-scale ($h_{\rm
      rot}$) relative to the half-light radius ($r_{50}$) as a
    function of stellar mass for galaxies in our sample. Galaxies are
    color-coded by the Sersic index as given in the key. The thick
    black line represents the median value as a function of mass.}
  \label{fig:size_scale}
\end{figure}

This section defines asymmetric drift  (\ref{sec:AD_SVDE}), and then steps through our measurement process of defining radial apertures and a common weighting of the data (\ref{sec:ap}), pressure corrections (\ref{sec:gas_cor}) to the gas tangential speed, and stellar population ages (\ref{sec:ages}) based on the \dnk~ spectral index.

\subsection{Asymmetric drift}
\label{sec:AD_SVDE}

Asymmetric drift ($V_a$) is defined as the lag between the tangential speed of the stars ($V_{\phi,*}$) and the circular speed of the potential: $V_a \equiv V_c - V_{\phi,*}$. We define an effective velocity dispersion $\sigma_{\rm AD}$ as the quadrature difference between the above quantities: 
\begin{equation}\label{eq:ad1}
\sigma_{\rm AD}^2 \equiv V_c^2 - V_{\phi,\ast}^2 = V_a (V_c + V_{\phi,\ast}).
\end{equation}
The two quantities $\sigma_{\rm AD}$ and $V_a$ are tightly correlated. In our data the correlation is characterized well by a slope of 1.5 between $\log V_a$ versus  $\log \sigma_{\rm AD}$, with a zeropoint of $V_a / \sigma_{\rm AD} \sim 1/4$ when $\sigma_{\rm AD} = 100$ \kms, i.e., where $V_a \sim 25$ \kms. Median values of $\sigma_{\rm AD}$ and $V_a$ for our sample are 70 and 16.5 \kms, respectively.

For the purpose of this work we adopt the ionized gas tangential velocity ($V_{\phi,g}$) as a proxy for $V_c$ and determine if any correction for pressure support is warranted. This correction is in the same sense as asymmetric drift, i.e., $V_{\phi,g} < V_c$, but arises for different physical reasons. For cool and cold gas, corrections for pressure support are minimal (of order a few \kms) and difficult to determine, and hence typically are ignored \citep[e.g.,][]{Swaters99}. For the warm ISM component, traced by ionized gas, \cite{Andersen06} have shown that the median H$\alpha$ line-width is $\sim18$ \kms\ in late-type spirals; this amounts to a negligible correction to $\sigma_{\rm AD}$ for $V_a$ as small as 10 to 20 \kms\ \citep[e.g.,][]{Westfall11}. This is consistent with the \kms-level agreement between HI and H$\alpha$ rotation curves for galaxies in the DiskMass Survey \citep{Bershady10}, as shown in \citet{Martinsson13-VI}. However, the MaNGA sample is more heterogeneous and displays a larger range of $\sigma_g$. Recent comparisons of CO and H$\alpha$ tangential velocities for a sample including MaNGA galaxies suggest modest corrections to the ionized galaxies velocities are needed in some cases where the gas dispersion is high \citep{Su22}. We develop and implement a correction for the ionized gas velocities in Section~\ref{sec:gas_cor}.

\begin{figure*}
  \centering
  \includegraphics[width=\linewidth]{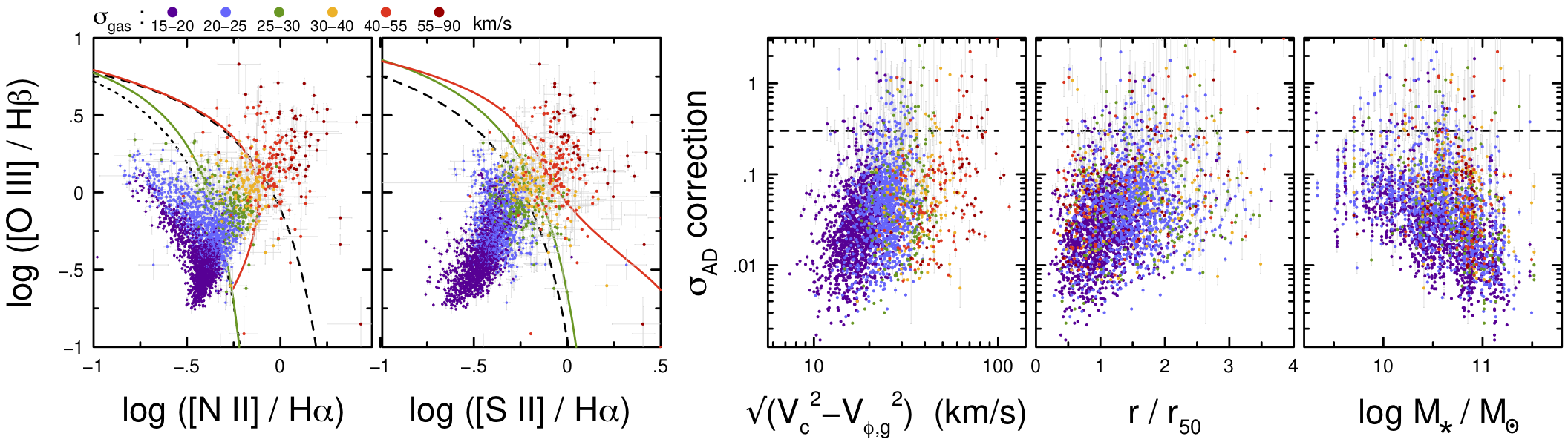} 
  \caption{Left panels: Emission-line flux ratio diagrams for H$\beta$, [O~III]$\lambda$5007, H$\alpha$, [N~II]$\lambda$6583, and [S~II]$\lambda\lambda$6716,6731 for measurements at all radii of our sample. Points are color-coded by estimated gas velocity dispersion, given in the key, and assigned according to line-ratios as described in the text. Black dotted and dashed lines represent demarcations of HII-like line-ratios from \citet{Kauffmann03} and \citet{Kewley01} respectively; solid green and red lines demark dynamically cold, intermediate and hot ionization regions as measured by \citet{Law21b}. Right panels: Fractional corrections to the measured asymmetric drift $\sigma_{\rm AD}$ for gas pressure-support as a function of the quadrature difference between the corrected and the measured gas tangential speed, radius (normalized by $r_{50}$) and stellar mass. Uncertainties include a full propagation of errors from gas and stellar velocities, line-ratios, and the population variance \citep[as measured by][]{Law21b} in the inferred $\sigma_g$ for a given set of line-ratios. The horizontal dashed line represents our threshold of a $<$30\% correction for inclusion in our remaining analysis.}
  \label{fig:pressure}
\end{figure*}

\subsection{Spaxel-based measurements in radial bins}
\label{sec:ap}

The inclined-disk fits used to select our sample also provide the rotation amplitude ($V_{\rm rot}$), the scale of the inner-rise of the rotation curve ($h_{\rm rot}$), and the kinematic position angle and inclination. The $\tanh$ rotation curve model is a rather simplistic description of the tangential motion, but it serves well to characterize the velocity gradients in the data, particularly at small radii. We measure the AD signal and other quantities directly from the data, examples of which are shown for AD in \cite{Shetty20}.

To deproject the observed spaxels to their relative radial and azimuthal positions on the galaxy we adopt the mean of the photometric and kinematic inclination, and the kinematic position angle. We use spaxels within $\pm40^\circ$ of the kinematic major axis (in the disk plane) to keep corrections for azimuthal projection to below 25\% while having a sufficiently broad wedge to sample an ample number of spaxels in each data-cube. The observed, line-of-sight radial velocities $V_{\rm los}$ for each spaxel are deprojected in the usual manner to derive the deprojected tangential speed $V = (V_{\rm los} - V_{\rm sys}) / (\cos\theta \sin i)$, where $V_{\rm sys}$ is the systemic velocity of each galaxy, and $\theta$ is the azimuthal angle in the disk plane for each spaxel.

We then compute weighted moments with $5\sigma$ clipping of the distributions of kinematic measurements made for individual spaxels, grouped in nested radial rings, 2 arcsec in width. The choice of width minimizes spatial covariance in data-cubes grided at $0.5\times0.5$ arcsec$^2$ from data sampled with 2 arcsec diameter fibers. An example of our binning and spaxel-based computations is given in Appendix~\ref{app:app} and Figure~\ref{fig:app_example}. We determine the error-weighted mean of $\sigma_{\rm AD}^2$ computed spaxel by spaxel, as defined in Equation~\ref{eq:ad1}. We stress this is not the same as taking the difference of the square of the mean stellar and gas velocities, i.e., $\langle V_{\phi,g}^2-V_{\phi,\ast}^2\rangle$ is not the same as $\langle V_{\phi,g}^2 \rangle-\langle V_{\phi,\ast}^2\rangle$. The former, in addition to being the correct formulation, has the practical advantage of enabling robust statistics and allowing for assessment of the spaxel-level distributions for measurements in the rare cases where the error-weighted mean of $\sigma_{\rm AD}^2$ is negative. These negative values are discarded from further analysis, and hence is a `survival' bias. However, this `survival bias' is inconsequential to the results of our analysis because the number of negative measurements are small (a few \% of the total).

We also compute the mean \dnk~ values as well as emission-line ratios for [O~III]/H$\beta$, [N~II]/H$\alpha$ and [S~II]/H$\alpha$ for the corresponding spaxels in each ring. We adopt \dnk~ as an age proxy, and present its conversion and calibration to a light-weighted mean age in Section~\ref{sec:ages}. We use the emission-line flux ratios in Section~\ref{sec:gas_cor} to estimate a correction to the gas velocity for pressure support. Since the $\sigma_{AD}$ kinematics are the paramount measure, often with the largest errors, and to avoid bias with different weightings, all mean values (specifically, emission-line ratios in Section~\ref{sec:gas_cor} and stellar population ages in Section~\ref{sec:ages}) are error-weighted using the inverse variance of $\sigma_{AD}$.

The total number of usable independent measurements from this sample is 2877 -- on average five to six measures per galaxy. We present results in five discrete radial bins between $0<r/r_{50}<3.5$ with upper limits in $r/r_{50}$ of 0.5, 1., 1.75, 2.5, and 3.5; the median values of $r/r_{50}$ for data in these bins are 0.4, 0.8, 1.3, 2.1 and 2.85. The number of measurements peaks in the second and third bins. Numbers drop in the inner bin because we exclude measurements for $r<4"$ based on the recommendations from \cite{Law21a} to avoid impact from beam-smearing. The median value of $r_{50}$ is 7.1 arcsec; at $0.6~r_{50}$ roughly half the sample has a radius $\sim$4 arcsec. In the outer two bins the survey design coverage limits the radial extent sampled. However, the distribution for apparent $r_{50}$ (in arcsec) has no trend in mass so the sampling in mass is uniform across radial bins.

The preponderance of measurements are at $r>h_{\rm rot}$, where the rotation curves are relatively flat. For our sample we find a median $h_{\rm rot}/r_{50}$ of 0.6. The distribution of this ratio as a function of stellar mass is shown in Figure~\ref{fig:size_scale}; the median drops smoothly from 0.7 to 0.45 in mass bins from $6.8\times10^9$ to $1.4\times10^{11}$ \Msol\, with a scatter (67\% CL) of 0.18 in any mass interval. A similar trend with rotation speed has been observed by \cite{Andersen2013}. The trend is in the anticipated sense that more massive galaxies have more concentrated mass profiles, as indicated by the increasing Sersic index with mass. This trend is, however, stronger for mass than for the Sersic index. Only the inner-most radial bin has some cases where $r \leq h_{\rm rot}$, but these are for lower-mass galaxies that have more slowly rising rotation curves so that in any event the radial velocity gradients within measurement annuli are small.

\subsection{Corrections for gas dispersion}
\label{sec:gas_cor}

\cite{Dalcanton10} have developed an elegant formalism for describing the anticipated lag between $V_c$ and $V_{\phi,g}$ due to pressure support in the gas otherwise moving in circular orbits in the galactic potential. \cite{Su22} show that this formalism does an excellent job of correcting H$\alpha$ tangential speeds to match that of the cold gas in disks. Referring to equation 11 of \cite{Dalcanton10}, this lag depends on the radial gradients of the gas velocity dispersion and density. In the context of pressure support in a multi-phase medium, the relevant gas density is the total gas density, while the gas dispersion is the radial component for the observed tracer (in our case H$\alpha$ from the ionized gas component, whose dispersion we assume is isotropic).

When both the gas dispersion and density profiles can be described as radial exponential profiles with scale lengths of $h_\sigma$ and $h_g$, respectively, then
\begin{equation}\label{eq:ad2}
V_c(r)^2 = V_{\phi,g}(r)^2 + \ r
\left[\frac{2h_g+h_\sigma}{h_\sigma h_g}\right] \sigma_g(r)^2.
\end{equation}
Since we do not have gas density profiles for individual galaxies \citep[see][]{Westfall14}, we leverage the results of \cite{Bigiel12} who find that $h_g \sim
0.61~r_{\mu25}$ in nearby spiral galaxies, with modest scatter ($r_{\mu25}$ is defined as the radius of the 25 mag arcsec$^{-2}$ $B$-band isophote). For an exponential stellar disk obeying Freeman's law \citep{Freeman70} with scale-length $h_d$ one finds $h_g \sim 1.88~h_d \sim 1.13~r_{50}$. (Using $r_{50}$ runs the risk of under-estimating $h_g$ for more bulge dominated systems where $r_{50}/ h_d < 1.67$, but the effect is minimal in practice.) In contrast, there is little evidence for significant radial gradients in the ionized gas velocity dispersion $\sigma_g$ outside of the inner regions where rotation curves steeply rise \citep{Andersen06,Martinsson13-VI}. Beam-smearing may be a significant contributor to the measured increase sometimes seen in $\sigma_g$ at small radii \citep{Law21a}, and our analysis here will work outside of this radial domain. It is therefore reasonable to assume that $h_g \ll h_\sigma$ so we adopt\footnote{Adopting $h_\sigma/h_g$ as large as 0.2 only changes the results of the correction by a few percent.}
\begin{equation}\label{eq:ad3}
 V_c(r)^2 = V_{\phi,g}(r)^2 + 0.88 \left[\frac{r}{r_{50}}\right]
 \sigma_g(r)^2.
\end{equation}

The amplitude of the correction to $\sigma_{\rm AD}$ can be estimated from equations~\ref{eq:ad1} and ~\ref{eq:ad3}.  In our analysis, we use the emission line-ratios to estimate our gas line-widths since almost all of our lines are well below the instrumental resolution of the data. An added advantage of this method is the minimization of the bias inherent in measuring line-widths at low-to-moderate S/N \citep{Law21a}. This estimation is based on the tight  correlation found by \cite{Law21b} between reliably-corrected MaNGA gas line-width measurements at high S/N ($>$50) and their line flux ratios of, e.g., [O~III]/H$\beta$ versus [N~II]/H$\alpha$. Specifically we use the average of the distance estimators d$_{N2}$ and d$_{S2}$ (defined in \cite{Law21b}, i.e., the distance in dex from the dynamically cold ridge-line in log([O~III]/H$\beta$) versus log([N~II]/H$\alpha$) or log([S~II]/H$\alpha$, respectively), and interpolate based on the mode and width of the histograms in Figure 3 of \cite{Law21b} for d$_{N2}$ to assign a characteristic value and uncertainty, respectively for the ionized gas line-width $\sigma_g$. 

The result of this inversion of the \cite{Law21b} correlation, applied to all of our measurements at all S/N, is shown in Figure~\ref{fig:pressure}, left-hand panels. Note that the preponderance of measurements are dynamically cold or HII-like (75\% based on d$_{N2}$ and 90\% based on d$_{S2}$, noting that the dynamically cold demarcation for [N~II]/H$\alpha$ is essentially identical to \cite{Kauffmann03}), and only 8\% are dynamically hot. The derived $\sigma_g$ values are qualitatively reasonable; quantitatively they agree with the measured $\sigma_g$ to within 1\% in the median, but with a 40\% dispersion dominated by the kinematic measurements errors and corrections. The advantage of the line flux-ratio inferred values is that $\sigma_g$ remains well defined at small line-widths where, in the presence of measurement error, the correction for the instrumental resolution to the measured kinematics sometimes lead to undefined values \citep[see][in the specific context of MaNGA data]{Westfall19,Law21a,Chattopadhyay24}.

The fractional value of this correction to $\sigma_{\rm AD}$ is shown in Figure~\ref{fig:pressure} as a function of radius and mass. The median correction is $<$~3\%; corrections for 67/90\% of the corrections are below 5/15\%.  We apply these corrections to derive $\sigma_{\rm AD,c}$. There are modest trends in the correction increasing with radius and decreasing with stellar mass, as expected from Equation~\ref{eq:ad3}, a relatively flat radial profile and mass-dependence for $\sigma_g$. We remove 124 measurements (4\%) with corrections $>30$\% from further analysis, but this does not preferentially bias against mass, radius, or ionization state. Four objects are removed from the sample, but these only had one or two initial measurements.

\subsection{Ages}
\label{sec:ages}

We adopt an empirical approach to estimate light-weighted mean ages for MaNGA spectra based on \dnk: Age estimates remain as closely linked to the data as possible and they are easily reproducible. Model estimates are used to calibrate our relation between \dnk~ and age.

Light- and mass-weighted mean ages have been estimated for MaNGA spectra by \citet[][Pipe3D]{Sanchez16a,Sanchez16b,Sanchez18}, \citet[][Firefly]{Goddard17,Parikh18} and \citet[][pPXF]{Lu2023} based on full-spectrum fitting to stellar populations synthesis (SPS) models. There are systematic differences in age estimates between models, the origin of which are not easily understood. Here we adopted the Pipe3D values for purposes of calibration, but again we emphasize that we do so in  a way that is transparent and easily convertible to other calibrations.

These SPS estimates rely on spectra that are spatially averaged to reach minimum S/N requirements. Unfortunately the spatial apertures do not match the geometric regions we require to measure kinematics and hence these data products cannot be used directly in this analysis.  This is secondary benefit of using \dnk~ that can be readily measured in the specific geometric regions of galaxies where we probe kinematics. 

\begin{figure}
  \centering
  \includegraphics[width=0.9\columnwidth]{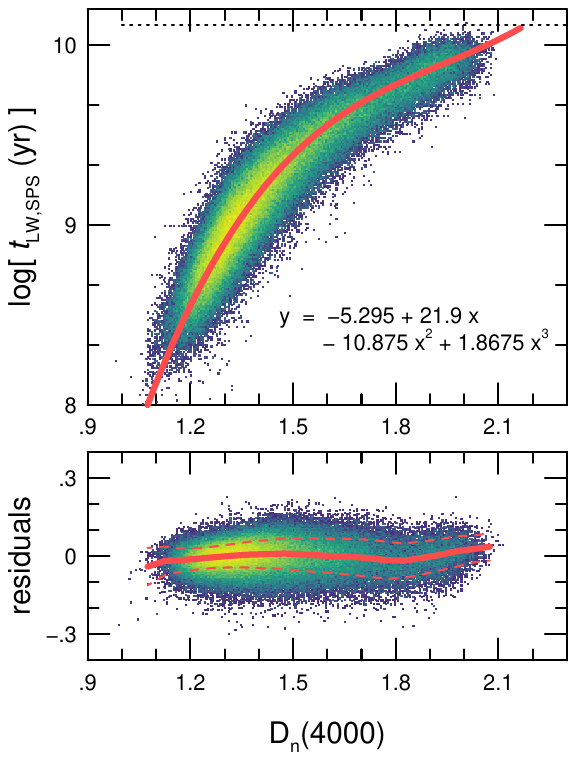}
  \caption{Light-weighted mean age ($t_{\rm LW,SPS}$) in the Johnson
    $V$ band derived from stellar population synthesis (SPS) in Pipe3D
    versus \dnk~ as defined in the DAP and measured here for 184447
    spectra in our sample. The red curve represents our adopted
    polynomial relation (coefficients listed) between y=$\log({\rm
      age})$ and x=\dnk. The bottom panel shows the orthogonal
    residuals. The median residual is given by the solid red curve,
    while $\pm$ the mean absolute deviation scaled by 1.48 to
    represent the standard deviation of a Gaussian distribution are
    rendered as dashed-red curves.}
  \label{fig:age_dn4000}
\end{figure}

Figure~\ref{fig:age_dn4000} shows the correlation of \dnk~ with
stellar population age. Data includes all spatial apertures defined by
Pipe3D for all galaxies in our sample with age estimates. We exclude
14\% of the apertures with the largest errors in \dnk~ ($>0.004$),
which appear to have uncertain age estimates based on their lack of
correlation with \dnk. Pipe3D does not measure \dnk~ so we measured
this quantity from the corresponding spectra, consistent with the
definition from the DAP. These apertures cover a broad range in
\dnk~ and age that spans our full data. The spaxels also sample the
full radial range of our data.

As shown in Appendix~\ref{app:age}, it is not possible to adopt a
simple model (SSP or single star-formation history) to match the
trends between age and \dnk. This illustrates the well-known need for
fitting SPS models to the data. We take advantage of this fitting, and
convert \dnk~ into an age using the third-order polynomial function
shown in Figure~\ref{fig:age_dn4000} with x=\dnk: $\log({\rm age}) =
-5.295 + 21.9 x -10.875 x^2 + 1.8675 x^3$.  This function traces the
measurements well with little ($<10$\%) systematic trends as a
function of \dnk. The formal random errors in age based on errors in
\dnk~ alone are much smaller than the width of the distribution. The
67\% CL in $\log({\rm age})$ is $\pm$0.13 dex (equivalent to a rms of
35\% in age), fairly constant with \dnk. The residuals illustrated in
Figure~\ref{fig:age_dn4000} are computed in the orthogonal direction
to the fit and are factor of 2.3 smaller than this. There are
residuals with metallicity and extinction (Appendix~\ref{app:age})
qualitatively consistent with expectations: When Pipe3D estimates
higher than average extinction or metallicity at a given \dnk~ the
corresponding Pipe3D age is lower. These residuals are consistent with
the observed distribution about our polynomial fit. Future work might
leverage the Balmer decrement and Lick indices to make suitable
corrections if systematics in the model calibration warrant.

The result of applying the conversion from \dnk~ to age is shown in
Figure~\ref{fig:ad_steps} at $0.8 \ r_{50}$. We refer to the
light-weighted mean age derived from \dnk~ as $t_{\rm LW}$. Galaxies
are color-coded by their total stellar mass. Not surprisingly, galaxies with
larger mass index tend to have larger \dnk~ values consistent with
older and more metal rich populations. These galaxies also have larger
$\sigma_{\rm AD}$.

%% file: analysis.tex
\section{Stratification Model}
\label{sec:stratification}

\subsection{Relating $\sigma_{\rm AD}$ to $\sigma_r$}
\label{sec:sAD_to_sr}

To relate our measure of AD to a stratification model requires
associating $\sigma_{\rm AD}$ with components of the stellar velocity
dispersion ellipsoid. Since the radial component of the stellar
velocity ellipsoid, $\sigma_r$, is typically the largest, and like
$\sigma_{\rm AD}$ probes in-plane motions, for simplicity we relate
$\sigma_{\rm AD}$ to $\sigma_r$. Starting with Equation~\ref{eq:app_ad3}
in Appendix~\ref{app:CBE} we adopt reasonable values for the ellipsoid
flattening and tilt, respectively, of $\sigma_z /
\sigma_r = 0.6$ \citep[e.g.,][]{Pinna18,Nitschai2021} and an intermediate value of $\lambda = 0.5$, which ranges from 0 to 1, to evaluate $\lambda[(\sigma_z/\sigma_r)^2-1]-1/2 = -0.82$ and arrive at:
\begin{equation}\label{eq:ad4}
\sigma_{\rm AD}^2 = \sigma_r^2 \left[ \frac{1}{2}\frac{\partial\ln
    V_\phi}{\partial\ln r} +
  \biggl(\frac{2b_n}{n_S}\biggr)\biggl(\frac{r}{r_{50}}\biggr) - 0.82
  \right]
\end{equation}
for a general Sersic profile of index $n_S$ where $b_n \sim 2n_S-1/3$ \citep[see, e.g.,][]{Graham05}. The term $\lambda[(\sigma_z/\sigma_r)^2-1]$ has a viable range between -1 and 0. The partial derivative for a $\tanh$ rotation-curve model, evaluated with equation~\ref{eq:drc}, is a subdominant term for $r/r_{50}>0.25$ (Figure~\ref{fig:ad_scale}).
Adopting the median values of $h_{\rm rot}/r_{50}$ = 0.6 (see
Figure~\ref{fig:pressure}) and an exponential light profile ($n_S=1$),
$\sigma_{\rm AD}/\sigma_r = 1.6^{+0.2}_{-0.1}$ at $r_{50}$, where
the nominal values assumes $\sigma_z /
\sigma_r=0.6$ and $\lambda=0.5$. The range
of values encapsulates $0.3<\sigma_z /
\sigma_r<1$ and $0<\lambda<1$.  A similar
computation for $1<n_S<5$ yields $\sigma_{\rm AD}/\sigma_r =
1.8^{+0.4}_{-0.1}$ at $r_{50}$. The weak dependence of this
$\sigma$-ratio on $n_S$ can be seen immediately by substituting in the
approximation for $b_n$ into the second term of Equation~\ref{eq:ad4}.
In the range $0.5<r/r_{50}<3$ for an exponential profile
$\sigma_{\rm AD}/\sigma_r$ can be well approximated as a linear
function of r given as $\sigma_{\rm AD}/\sigma_r = 0.85 +
0.75(r/r_{50})$. This approximation is useful for relating
$\sigma_{\rm AD}$ to components of the stellar velocity ellipsoid, but
we use equation~\ref{eq:ad4} for our analysis. We also apply the
measured $h_{\rm rot}$ and $r_{50}$ values for our sample. In the
context of the model there is at most 10\% systematic uncertainty in
this scaling for a plausible range of $\lambda$ and $\sigma_z / \sigma_r = 0.6$ at a
given radius.

\subsection{Stratification model parameterization}
\label{sec:strat_mod}

The stratification of disk stars in phase-space may be either a relic
of disk-settling, i.e., gas-cooling \citep{Brook04,Bournaud09}, or the
result of a secular dynamical process of disk heating
\citep{Spitzer51,Spitzer53}. Possibly it is a combination of both
\citep{Bird13,Martig12,Martig14a,Martig14b}. Here we wish merely to
have a simple model that describes the present-day stratification that
(a) reasonably describes what we see in the Milky Way's solar
neighborhood; (b) can be generalized to other radii and galaxies on
basic astrophysical grounds; and (c) this generalization can be tied
as directly as possible to observable quantities.

For this reason, we turn to the long-standing disk-heating model proposed by several authors \citep{Wielen77,Binney00,Aumer09,Aumer16a} who posed that for isotropic scattering, independent of stellar orbit, $ d(V^2)/dt = D~t$ where $D$ is the diffusion coefficient inversely proportional to V. From this one finds that 
\begin{equation}\label{eq:sot}
\sigma(t) = \sigma_0 ( 1 + t/\tau)^b, 
\end{equation}
with $b=1/3$ and parameters for the solar cylinder of $\sigma_0= 6$ \kms\ and $\tau = 0.05$ Gyr. 

These values for $\sigma_0$ and $\tau$ are consistent with dispersions for the cool atomic and molecular gas layer and the dynamical time-scale of the disk at the solar circle, respectively. We use the latter fact to make a general model applicable at all radii by adopting
\begin{equation}\label{eq:heatmod}
\tau(r) \equiv t_{\rm dyn}(r) = (\pi/2) \ r / V_c(r),
\end{equation}
i.e., $t_{\rm dyn}(r)$ is 1/4 of the orbital period at radius $r$.\footnote{In general the epicyclic frequency $\kappa^2= 2 (V_c/r)(V_c/r + \partial V_c/\partial r)$ for nearly circular orbits in an axisymmetric potential \citep{Binney08}. Hence in the radial range where the rotation curve is flat we can relate $t_{\rm dyn}(r)  = (\pi/\sqrt{2}) \ \kappa(r)^{-1}$.} 

In the case where the perturbers are highly concentrated to the disk mid-plane (e.g., giant molecular clouds), it might well be argued that the vertical oscillation period about the disk mid-plane, $T_z(r)$, would be better suited to equate with $\tau$ in Equations~\ref{eq:sot} and \ref{eq:heatmod}. For nearly circular orbits in an axisymmetric potential the vertical oscillation period, $\nu$, can be found from the second derivative of the potential with respect to height. This yields $T_z(r) = 2\pi/\nu = \sqrt{2\pi/G\rho_0(r)}$, where $\rho_0(r)$ is the disk mid-plane density at radius $r$ \citep{Binney08}. 

As it so happens at the solar circle, $T_z(r)$ and $t_{\rm dyn}(r)$ are nearly the same even though their scalings with radius are different. If we assumes a constant mass scale-height with radius, and that dynamical mass surface-density scales with surface-brightness, then for the MW (using parameters adopted in Section~\ref{sec:MW}) we find $t_{\rm dyn}/T_z = 1\pm0.1$, i.e., unity to within 10\%, in the range $0.7<r/r_{50}<2.7$. This range contains 83\% of our measurements. In our smallest and largest radial bins the difference between the two time-scales is $<35$\%. The uncertainties in estimating mass scale-heights and mass surface-densities (particularly when atomic and molecular gas mass surface-densities are not available) to determine $\rho_0$ and hence $T_z$ are likely larger than the differences between $t_{\rm dyn}(r)$ and the actual $T_z(r)$ values at any given radius of interest. In contrast, the circular speed is reliably measured in our data. For these reasons we opt to retain $t_{\rm dyn}(r)$ rather than $T_z$ for our reference time-scale $\tau(r)$.

\begin{figure*}
  \centering
 \includegraphics[width=\textwidth]{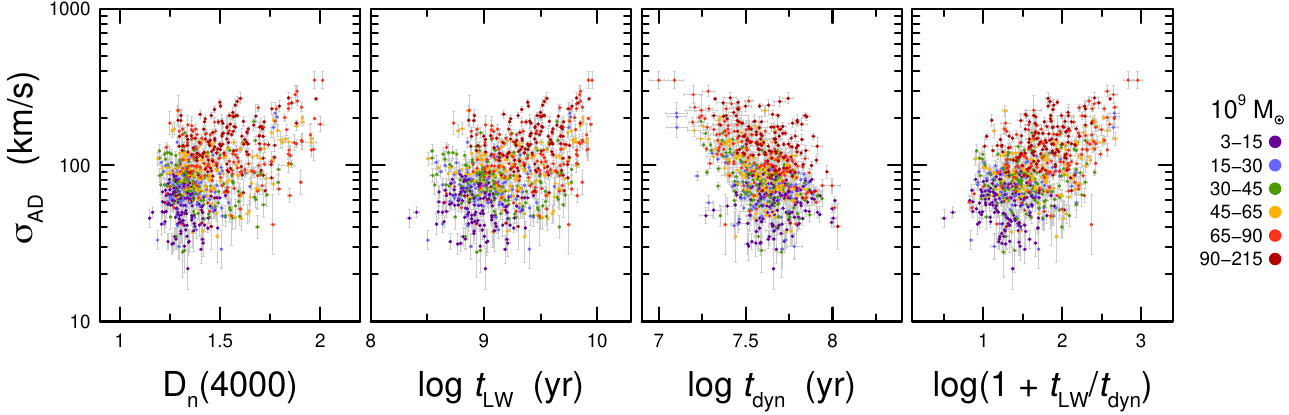}
  \caption{Asymmetric drift signal, $\sigma_{\rm AD}$ at a median radius of 0.8 $r_{50}$ versus (left to right): \dnk, light-weighted mean age ($t_{\rm LW}$) inferred from \dnk, local dynamical time ($t_{\rm dyn}$, see text), and the ratio of these two time scales. Each point represents a galaxy measurement in the radial range from $0.5<r/r_{50}<1$, color-coded by stellar mass as given in the key.}
  \label{fig:ad_steps}
\end{figure*}

\cite{Wielen77} pointed out that another reasonable assumption is for the diffusion coefficient $D$ to be constant, yielding a power-law index $b = 0.5$ rather 0.33. Numerous theoretical predictions and observational analyses of stellar ages and kinematics in the MW solar neighborhood yield $0.2<b<0.6$; these are discussed in our introduction with the theoretical and observational literature also well summarized by \cite{Kumamoto17} and \cite{Hanninen02}, respectively. It is also clear from these studies that the values of $b$ and $\sigma_0$ depend on the component of the velocity ellipsoid: For the MW solar neighborhood the vertical index ($b_z$) tends to be larger than the radial $b_r$, but with lower initial amplitude \citep[$\sigma_{z,0}<\sigma_{r,0}$;][]{Hanninen02, Holmberg09, Aumer09, Sharma14, Aumer16a}.

Consequently for our analysis we specifically use our conversion
$\sigma_{\rm AD}$ to $\sigma_r$, and in fitting trends of $\sigma_r$
with age and radius, we allow both the amplitude ($\sigma_{r,0}$) and index 
($b_r$) to be model parameters:
\begin{equation}\label{eq:sor}
\sigma_r(t) = \sigma_{r,0} ( 1 + t/t_{\rm dyn})^{b_r},
\end{equation}
where $t_{\rm dyn}$ is defined by Equation~\ref{eq:heatmod}. We make no assumptions about $b_r$, $\sigma_{r,0}$, their variation between galaxies, or their trends with radius. We stress that this {\it stratification} model says nothing about time's arrow and simply is a convenient way to predict the relation between stellar population age and kinematics (i.e., their AVR) in today's galaxies {\it if} their stratification process is similar to that of the MW. In a large sample of galaxies, we might expect the AVR in some cases is dominated by one or more impulsive events \textit{ex situ} to the disk, e.g., mergers; in these cases our stratification model may be inadequate to describe well the data. This is something we can test. We also adopt the light-weighted mean age (\tlw) for $t$, which carries  an implicit assumption that the light-weighted age as determined from SPS (continuum shape and absorption-line equivalent widths) corresponds to the characteristic asymmetric drift signal measured from the stellar kinematics. We address low-level systematics inherent in this assumption below.

\section{Analysis}
\label{sec:analysis}

\subsection{Systematics between light-weighted ages and kinematics}
\label{sec:lwa_sys}

While we measure light-weighted ages and kinematics from full-spectrum fitting over the same range in wavelength, the effective light-weighting for stellar age and stellar kinematics (here, velocities) are not necessarily identical. This might come about, for example, because the relevant line-strengths that contain the kinematic signal do not scale with age in the same way as the continuum level. Further, the stellar population ages are, by definition, referenced specifically to some wavelength or wavelength range (we chose the $V$-band to be near the mid-point in the spectral range), and, unlike the kinematics, do not depend in detail on the wavelength distribution of relative line-strengths. These differences may lead to subtle systematic mismatches between characteristic velocities and ages. We addressed this by creating mock spectra with star-formation histories and rates of dynamical stratification that appear to span what is observed in our data, as described in Appendix~\ref{app:lwa_sys}.

We find that in the presence of dynamical stratification there are systematic differences between recovered velocities and velocities expected from $\sigma_r(t)$ given the recovered light-weighted age: pPXF tends to overestimate ages and velocities (and hence underestimate $\sigma_{\rm AD}$). The amplitude of the systematics depends primarily on $V_c$ and \dnk, peaking at the lower values of \dnk\ $\sim1.3$ for the fastest rotation speeds. While the correction can be as much as 30\%, the mean correction is 11\% at $r_{50}$, increasing on average with radius from 8\% to 14\%  between $0.25<r/r_{50}<3$. The typical uncertainty in this correction increases from 2\% to 4\% over the same radial range, as illustrated in the right-hand map of Figure~\ref{fig:lwa_sys}. In summary, the corrections are quite modest and well determined. They are applied to the data, based on the $V_c$ and \dnk values of each radial bin in each galaxy, in all subsequent analysis.

\begin{figure*}
  \centering
 \includegraphics[width=\textwidth]{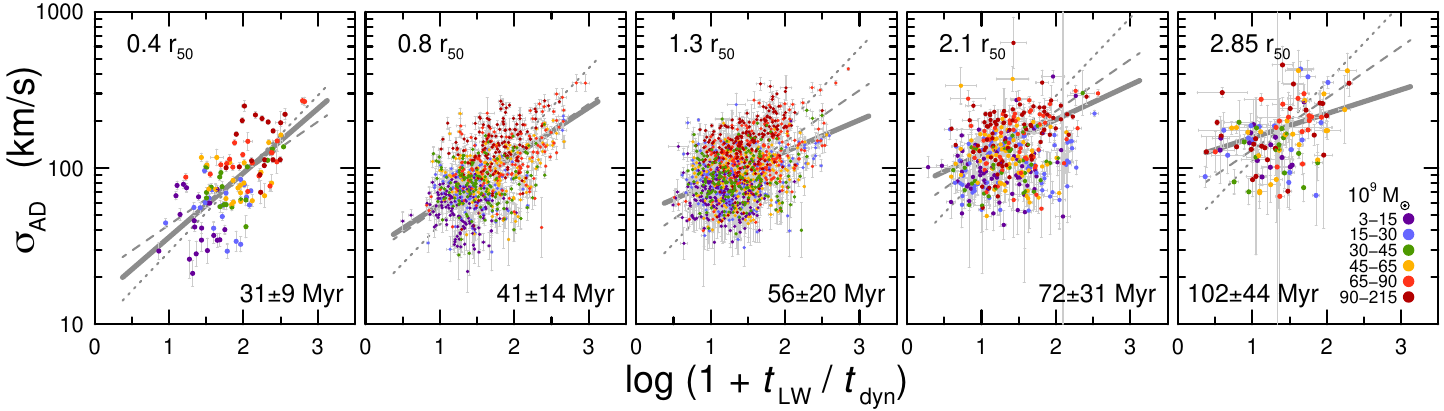}
  \caption{Asymmetric drift signal, $\sigma_{\rm AD}$, versus the ratio of stellar population light-weighted age to dynamical time for five radial bins scaled by $r_{50}$ given in the top left of each panel; the median and 67\% range of dynamical times are given at the bottom of each panel. Points represent measurements from individual galaxies at a particular radius, color-coded by stellar mass as given in the key. Solid grey curves represent the best-fitting stratification model to the ensemble of galaxies at each radius. Dashed and dotted curves correspond to stratification models with $b_r$ fixed to 0.33 and 0.5, respectively.}
  \label{fig:ad_rbin}
\end{figure*}

\subsection{Stratification model applied to MaNGA data}
\label{sec:application}

Figure~\ref{fig:ad_steps} shows the trends of $\sigma_{\rm AD}$ at 0.8 $r_{50}$ with \dnk, light-weighted mean age, local dynamical time, and the ratio of light-weighted mean age to the local dynamical time. Notable are the opposite trends of the two time-scales with $\sigma_{\rm AD}$ and stellar mass. On the one hand stellar age correlates with total mass, while at a given radial scale the increase in the rotation speed in more massive galaxies decreases the dynamical time. These two factors are both at play in the correlation of $\sigma_{\rm AD}$ with $\log( 1 + t_{\rm LW}/t_{\rm dyn} )$ that we use to constrain the stratification model. We measure $\tau_{\rm dyn}$ directly from the data, finding values of $41\pm14$ Myr at $r/r_{50} = 0.8$, with the range enclosing 67\% of the sample in the radial bin.

\subsubsection{Fitting to galaxy ensembles}
\label{sec:ensemble}

There are two ways we are able to fit the stratification model to the data. The first is to aggregate measurements across all galaxies. This has the advantage of maximizing the number of measurements, and therefore is conducive to breaking down the data by radius, total galaxy mass, or both. The potential pitfall is the inherent assumption that the stratification parameters, even at a given radius and galaxy mass, are the same for different galaxies. Variations in both the amplitude ($\sigma_{r,0}$) and heating index ($b_r$) between galaxies may wash out the correlation between $\log( 1 + t_{\rm LW}/t_{\rm dyn} )$ and $\sigma_{\rm AD}$, possibly yielding best-fitting values for $b_r$ that are systematically too low and values for $\sigma_{r,0}$ that are too high.

Figure~\ref{fig:ad_rbin} shows the result of applying the stratification model to our data as an ensemble.  The results are shown to fits done in different radial bins, adopting an exponential light profile for the disk. Median dynamical times and ranges are given for each radial bin. We use the scaling from $\sigma_r$ to $\sigma_{\rm AD}$ given in Section~\ref{sec:AD_SVDE} (equation~\ref{eq:ad4}), and compute $\sigma_r(t)$ at each radius according to equation~\ref{eq:sor}. We then find $\sigma_{r,0}$ and $b_r$ at each radius that minimizes $\chi^2$ defined using the differences between the measured and predicted asymmetric drift ($\sigma_{\rm   AD}$); we define $\chi^2$ as the orthogonal distance to the model line defined by Equation~\ref{eq:sor} \citep[essentially an orthogonal least-squares formulation in the presence of heteroscedastic errors in both variables][]{Akritas96}. A minimum $\chi^2$ value is found numerically, by brute force, with the help of a zoom-in function for added precision once the global minimum is found. This `best fit' model (in a population sense) is represented as the thick grey line in Figures~\ref{fig:ad_rbin}. Alternatively, by fixing $b_r$ to historically preferred values for the MW of 0.33 and 0.5, we can repeat the exercise to find the corresponding best values of $\sigma_{r,0}$ at each radius. These are shown as dashed and dotted lines in Figures~\ref{fig:ad_rbin}. In a $\chi^2$ sense these fixed-$b_r$ models do not do as well.

\subsubsection{Fitting to individual galaxies}
\label{sec:indy}

We also fit the stratification model in a second way, namely to data for individual galaxies. We refer to this as the `indy' method. Again, we adopt an exponential light profile for the disk. In this case there are significant limitations in the number of measurements per fit as well as the dynamic range in both $\log( 1 + t_{\rm LW}/t_{\rm dyn} )$ and $\sigma_{\rm AD}$. Conveniently, most galaxies exhibit significant radial population gradients.

In contrast with the ensemble method, the inherent assumptions with the indy method are that the stratification parameters are the same across radius, while assumptions about the impact of variations in the merger history, or satellite and halo populations are not an issue as they are in the ensemble method.

In practice, we find there are diminishing returns to deriving reliable stratification parameters when the range in the sampled time-scale ($\Delta\log( 1 + t_{\rm LW}/t_{\rm dyn} )$) is within a factor of a few of the combined errors in $x=\log( 1 + t_{\rm LW}/t_{\rm dyn} )$ and $y=\log  \sigma_{\rm AD}$. Defining the latter two errors as $e_x$ and $e_y$, and the mean of the combined errors across the measurements for an individual galaxy as $\langle (e_x^2+e_y^2)^{1/2} \rangle$, in our subsequent analysis we restrict our consideration to stratification parameters derived from indy fitting to galaxies with 
\begin{equation}\label{eq:snr}
\Delta\log( 1 + t_{\rm LW}/t_{\rm dyn} ) / \langle (e_x^2+e_y^2)^{1/2}\rangle > 8. 
\end{equation}
We also require there to be at least 4 radial measurements per galaxy.\footnote{This eliminates all 14 sample galaxies observed with the smallest 19-fiber IFU, and roughly half and 20\% of sample galaxies observed with the 37-and 61- fiber IFUs, respectively. We do not find systematic differences in the stratification parameters of the remaining galaxies as a function of IFU size.} This reduce the available sample to 221 galaxies (145 unbarred). The results of these fits, unexpectedly, yield noisy stratification parameter estimates for any given galaxy (see Section~\ref{sec:discussion} and associated Figures below), but their statistical averages, as discussed below, provide useful constraints on these parameters.

\begin{figure*}
\centering
   \includegraphics[width=0.9\textwidth]{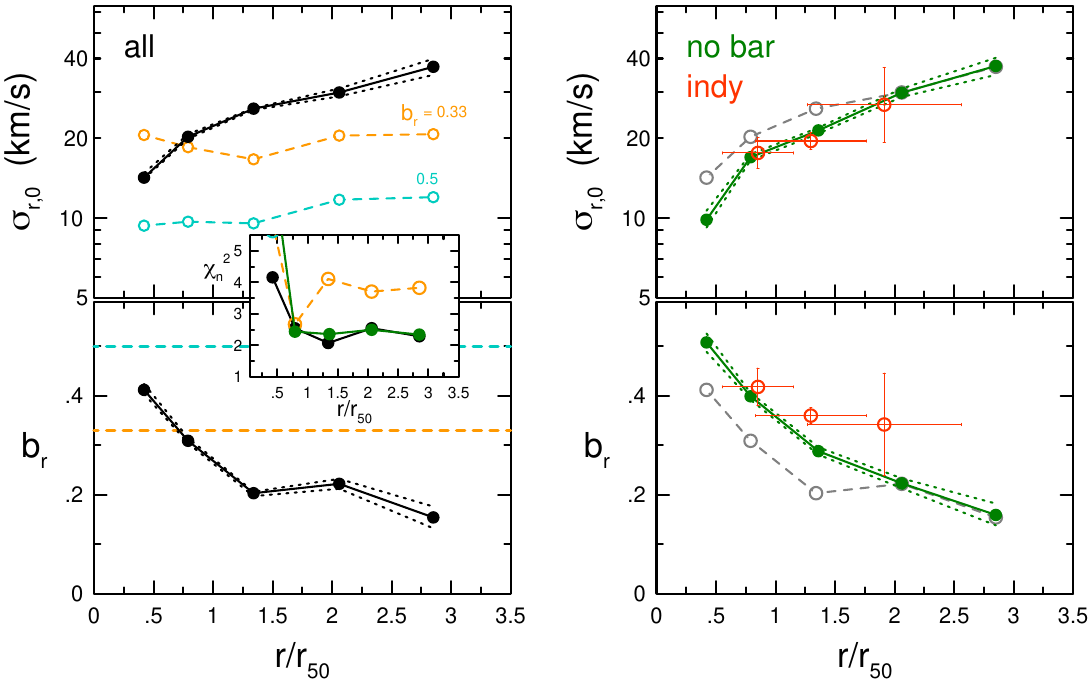}
   \caption{Stratification parameters $\sigma_{r,0}$ and $b_r$ versus radius scaled to $r_{50}$. Radial bins correspond to Figure~\ref{fig:ad_rbin}. \textit{Left:} Best-fitting models fit simultaneously to data from all galaxies in a given radial bin. The black curve represents when both parameters free (dotted lines give 67\% CL uncertainties on combined parameters). Best fitting values for $\sigma_{r,0}$ for $b_r$ fixed to values of 0.33 and 0.5 are shown as orange and teal lines and dots, respectively. The inset plots the trend of reduced-$\chi^2$, $\chi^2_\nu$, for these different cases ($\chi^2_\nu$ for $b_r=0.5$ is off scale), as well as the unbarred sample shown in the right panel. \textit{Right:} Best-fitting models to all unbarred galaxies are shown in green (grey repeats results for all galaxies from the left panel). Red data points are averages for fits to individual galaxies, binned by the mean radius of the measurements for a given galaxy, as described in the text.}
   \label{fig:tdyn_adm_best}
\end{figure*}

\subsection{Radial trends}
\label{sec:rad_trend}

The immediate impression from Figure~\ref{fig:ad_rbin} is that there is a significant trend with radius of increasing overall amplitude of $\sigma_{\rm AD}$ at a given $\log( 1 + t_{\rm LW}/t_{\rm dyn} )$, as well as a \textit{flattening} of the correlation between these two quantities. The derived model parameters $\sigma_{r,0}$ and $b_r$ are plotted in Figure~\ref{fig:tdyn_adm_best} (left-hand panel) as a function of radius adopting an exponential light profile for the disk (equation~\ref{eq:ad4}).\footnote{The results adopting the generalized Sersic profile and Sersic indices for each galaxy yielded nearly identical results, so we do not report them here.} We clip the data at 2.5$\sigma$ about the best fit; reducing the clipping does not change the qualitative trend, but the results are considerably noisier. Confidence intervals (ensemble method) and weighted errors (indy method) are computed on the clipped sample. The stratification amplitude $\sigma_{r,0}$ rises with with radius, between 15 and 40 \kms. The stratification index $b_r$ also declines with radius in the range $0.4<b_r<0.2$. The results for fixed $b_r=0.33$ and 0.5 serve to depress best-fitting values for $\sigma_{r,0}$ and also keep these values nearly constant with radius; this simply illustrates well the covariance between the two stratification parameters.

\begin{figure*}
\centering
   \includegraphics[width=\textwidth]{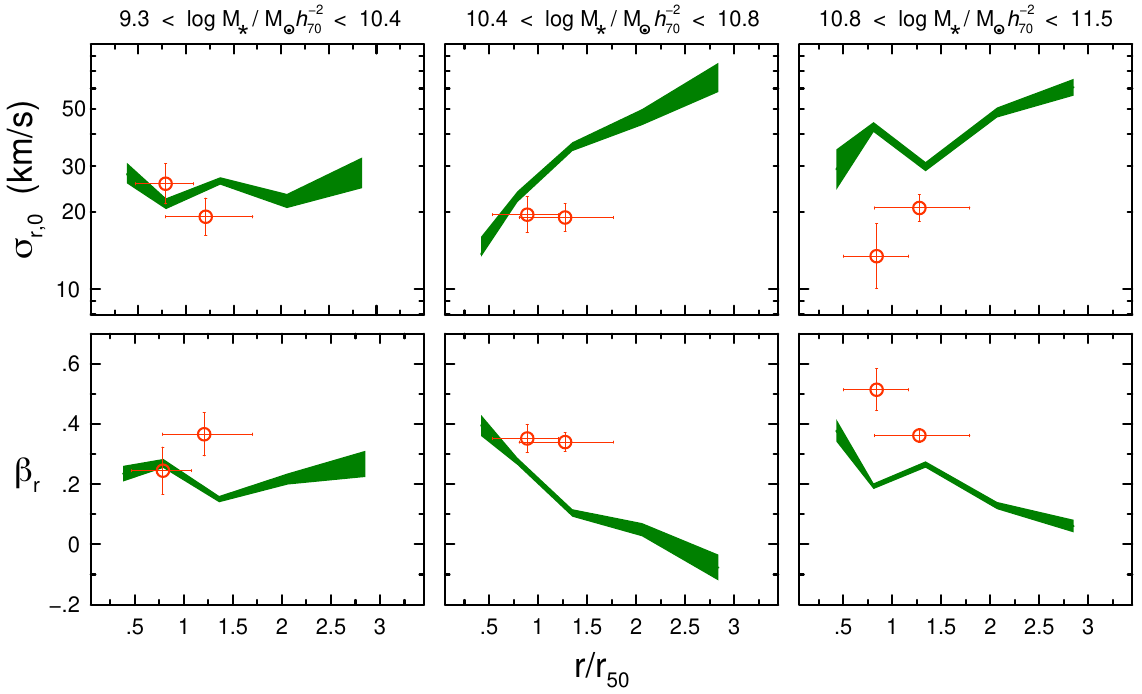}
   \caption{Stratification parameters $\sigma_{r,0}$ and $b_r$ versus radius scaled to $r_{50}$ for three bins in stellar mass (solar units) for \textit{un}barred galaxies. Radial bins correspond to those in Figures~\ref{fig:ad_rbin} and \ref{fig:tdyn_adm_best}. Best parameters and 67\% CL from fitting the model to the galaxy ensemble are shown as green polygons. Open points in red represent weighted averages of the best model parameters for fits to individual galaxies, binned by the mean radius of the measurements for a given galaxy, as described in the text. Error bars represent the error in the weighted mean (y), and the width of the radial bin encompassing 67\% of the measurements (x).}
   \label{fig:tdyn_mass_radius}
\end{figure*}

\subsubsection{Barred and unbarred samples}
\label{sec:bars}

Barred galaxies comprise $\sim$23\% of our sample. It is reasonable to expect that in radial regions where bars are present, our formulation for inferring velocity dispersion from asymmetric drift may break down. Bars are due to resonances that perturb orbits far from circular motion and can lead to gas shocks that strongly differentiate between (collisional) gas and (collisionless) stellar radial and tangential motions. 

In the right-hand panel of Figure~\ref{fig:tdyn_adm_best}, we recompute the best-fitting stratification parameters as a function of radius for the ensemble method, this time excluding galaxies with visually identified bars (green curves). The result leads to a smoother and steeper trend in $b_r$ with radius, peaking at 0.5. Commensurately, the amplitude of $\sigma_{r,0}$ decreases. There is no change outside of $r/r_{50}=1.75$, as one might expect since bars do not modulate orbits at large radii. Trends with radius in the stratification parameters fit only to barred galaxies are not plotted; we find they are very erratic, leading, for example, to strongly negative values of $b_r$ at intermediate radius. For this reason we focus our remaining analysis on the unbarred sample.

\subsubsection{Ensemble versus individual galaxy fitting methods}
\label{sec:ensemble_vs_indy}

The right-hand panel of Figure~\ref{fig:tdyn_adm_best} also compares results for unbarred galaxies for the ensemble and indy methods (the latter are shown as red points). For the indy method we compute a weighted average of the stratification parameters for individual galaxies in radial bins according to the mean radius of the measurements for that galaxy. As a result, the radial bins in the indy-method averages are over a broader range of radii than for the ensemble-method. Given the MaNGA target selection \citep{Wake17}, there is also a slight increase in the mean mass at larger mean radii; the effect is less than 0.1 dex, but primarily there are no galaxies below $10^{10}$ M$_\odot$ in the largest indy radial bin. The agreement in $\sigma_{r,0}$ is remarkably good. The trend with $b_r$ is somewhat flatter for the indy-method results, but still consistent within the uncertainties with the ensemble method; the flattening may be due to the wider radial coverage in the indy method. Overall, this comparison appears to indicate that any systematics due to the assumptions described in Section~\ref{sec:application} are not particularly strong.

\begin{figure}
\centering
   \includegraphics[width=0.9\columnwidth]{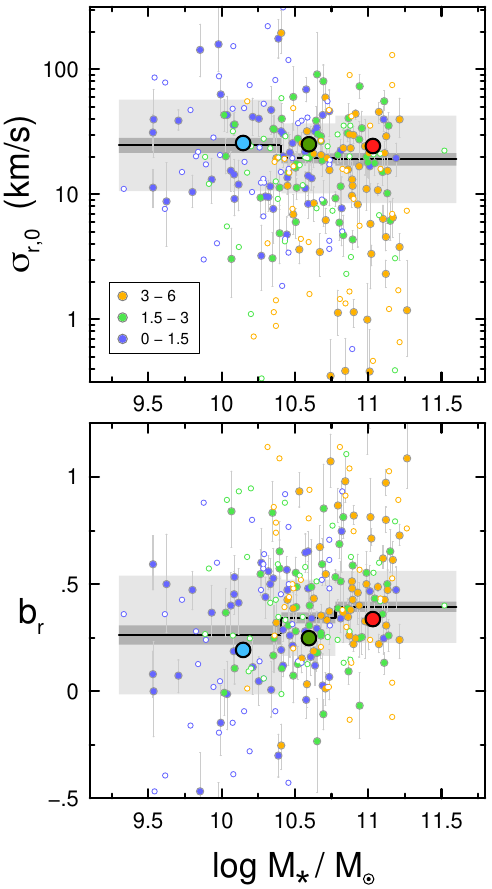}
   \caption{Stratification parameters $\sigma_{r,0}$ and
     $b_r$ versus galaxy total stellar mass for all radii combined. Small circles represent results for the indy method (individual galaxies), color-coded by their Sersic index as given in the key. Open circles represent galaxies that do not have sufficient dynamic range in $t_{\rm LW}/t_{\rm dyn}$ (Equation~\ref{eq:snr}). Light grey, dark grey and black lines represent the weighted standard deviation, error in the mean and mean for the indy method measurements. Large, filled circles represent the best fit values for the ensemble method. }
   \label{fig:tdyn_mass}
\end{figure}

\subsection{Correlation with galaxy stellar mass}
\label{sec:strat_var}

Another immediate impression from Figure~\ref{fig:ad_rbin} is that at any given radius there is a gradient in total galaxy mass along the trend-line between the $\sigma_{\rm AD}$ and $\log( 1 + t_{\rm LW}/t_{\rm dyn} )$. This differentiation weakens at larger radius where the slope also appears to lessen. It is reasonable to wonder if, or to what extent a mass-dependence in both quantities drives the correlation between these two parameters. That is to say, it is conceivable that there exist \textit{primary} causal links between mass and both mean stellar population age (as a reflection of trends in star-formation history with halo mass) and stellar velocity dispersion (a proxy for mass in dispersion-dominated systems), the result of which is an \textit{apparent} correlation between $\sigma_{\rm AD}$ and $\log( 1 + t_{\rm LW}/t_{\rm dyn} )$ that is not driven by a stratification process. Even in systems that are not dispersion-dominated (such as galaxy disks), it is reasonable to expect that more massive galaxies will tend to have more massive disks with larger velocity dispersions, and hence larger $\sigma_{\rm AD}$.

Alternatively, we suggest this mass-gradient manifests in the data as a reflection of a general age-velocity relationship in stellar populations where (i) more massive galaxies tend to have more evolved (older and more metal rich) stellar populations (i.e., the trends in the CMD); and (ii) within galaxies the inner regions tend to have older stellar populations (so-called inside out growth). Equivalently: late-type galaxies tend to be less massive and have smaller bulges (and more modest population gradients), while early-type galaxies tend to be more massive and have larger bulges (and more pronounced populations gradients transitioning from bulge to disk). These trends reflect the distribution of, e.g., \dnk, within any given galaxy. The correlation between  $\sigma_{\rm AD}$ and $\log( 1 + t_{\rm LW}/t_{\rm dyn} )$, in this picture, represents a general age-velocity relationship in stellar populations with all disk galaxies. The fact that there is general agreement between the stratification parameters in the ensemble and indy methods (Figure~\ref{fig:tdyn_adm_best}) supports this picture. 

To further demonstrate that this correlation is not driven entirely by the correlation of mean stellar population age with galaxy stellar mass, we can also divide our sample by mass into three equal bins by number of galaxies (and roughly equal number of individual measurements) and repeat the fitting exercise. The mass boundaries are $\log {\rm M}_\star / {\rm M}_\odot = 10.4$ and 10.8. The median mass in each bin is 10.2, 10.6 and 11.0 in the log, with bins widths of 0.9, 0.4, and 0.5 dex\footnote{In the lower- and upper-mass bin width we discount the full bin-width required to include one galaxy in each bin at the extra of the mass distribution that do not have indy-method fits.}. The results, as a function of both mass and radius, are illustrated in Figure~\ref{fig:tdyn_mass_radius} for both ensemble and indy methods. Again, there are clearly non-zero values for $b_r$ in all mass bins and at most radii, further supporting a stratification model driving a general age-velocity relationship in stellar populations. It is also interesting to note that the stratification parameters at $r=r_{50}$ are very similar in all mass intervals, despite the different radial trends. This may suggest a self-similar stratification mechanism at this radius across galaxy masses.

\subsubsection{Stratification rate versus stellar mass}
\label{sec:strat_var_mass}

While the trends in the stratification parameters with radius appear somewhat more erratic when divided by mass (due to smaller sub-samples), the general trends with radius in $\sigma_{r,0}$ and $b_r$ seen for the full sample are evident for the intermediate and higher-mass sub-samples based on results for the ensemble method: the stratification index ($b_r$) is lower at larger radii. In contrast, there is little evidence for a radial gradient in the stratification parameters for $\log {\rm M}_\star / {\rm M}_\odot < 10.4$. This suggests the stratification process may be different in lower-mass galaxies. The results for the radial trends from the indy method are more difficult to assess given their limited dynamic range in mean radius, but the results do suggest that the stratification index ($b_r$) \textit{increases} with galaxy mass. 

The mass-dependence of $b_r$ is seen more clearly when we recompute the marginal trend in mass alone  by fitting to all radii in each mass bin (ensemble method) or combining all galaxies of a given mass for the indy method. The results shown in Figure~\ref{fig:tdyn_mass} display a clear trend of the stratification parameters with mass for \textit{both} methods. The results in these two Figures \textit{demonstrate that the relation between $\sigma_{\rm AD}$ and $\log( 1 + t_{\rm LW}/t_{\rm dyn} )$ is driven by a stratification process whose index ($b_r$) increases with mass.}

\subsection{Discussion}
\label{sec:discussion}

Overall the stratification index ($b_r$) appears to be lower for the ensemble method than the indy method. Our preference here is for the indy method values: Because the effective radial range of the indy measurements is modest, systematics are also likely modest. The ensemble method, as already noted, may tend to bias toward lower $b_r$ and higher $\sigma_{r,0}$ if the population exhibits a significant range of stratification. We find no appreciable change in minimum reduced $\chi^2$ ($\chi_\nu^2$) or sample rejection fraction (with iterative clipping) in the ensemble method when dividing the sample by mass. This suggests additional astrophysical scatter or the limited efficacy of the simple model to explain age-velocity relations in the galaxy population. 

We explored if the dependence on stratification parameters was better defined by the Sersic index (as a measure of stellar mass concentration). Figure~\ref{fig:tdyn_mass} also serves to show that trends in the stratification parameters with mass are mimicked by trends of the Sersic index with mass. Given the range of the stratification parameters at a given mass and the strong correlation of Sersic index with mass, we have been unable to find any additional significant dependence on the Sersic index.

\begin{figure}
\centering
   \includegraphics[width=0.9\columnwidth]{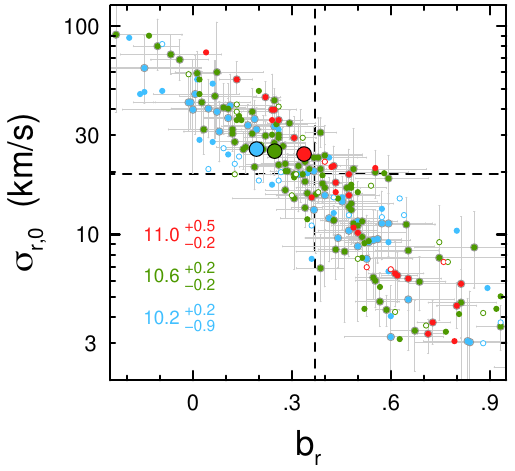}
   \caption{Covariance in stratification parameters $\sigma_{r,0}$ and $b_r$. Symbols are as in Figure~\ref{fig:tdyn_mass}, except color-coded by total stellar mass as given in the key ($\log {\rm M}_\star / {\rm M}_\odot$). Dashed lines correspond to the mean parameter values for the indy method averaged over all mass. }
   \label{fig:ad_covar}
\end{figure}

As one might anticipate from Equation~\ref{eq:sor}, and as noted above (Figures~\ref{fig:tdyn_adm_best}) through ~\ref{fig:tdyn_mass}, there is a strong covariance between $\sigma_{r,0}$ and $b_r$. This is best seen in the stratification parameter values determined for individual galaxies, as shown in Figure~\ref{fig:ad_covar}. The slope of the correlation between $\log\sigma_{r,0}$ and $b_r$ is well described by the characteristic value of $-\log(1+t/t_{\rm dyn})=-1.5$, as seen from inspection of Figure~\ref{fig:ad_rbin}, and as it should be; the correlation is slightly steeper(shallower) for the higher(lower) mass galaxies, also consistent with the trends in the $\log(1+t/t_{\rm dyn})$ distributions as a function of mass as seen in this figure.

%% file: local_group.tex
\section{Dynamical stratification in the Local Group}
\label{sec:LG}

\subsection{Local Group age--velocity dispersion relations}
\label{sec:LG_AVR}

There can now be found in the literature measurements of stellar
velocity dispersion and age for Local Group galaxies to directly
compare our results. Here we limit our comparison to three galaxies
with comparable range of mass as our MaNGA sample, namely (in order of
decreasing mass) M31, the Milky Way and M33.  We adopt total stellar
masses of $1.04\pm0.05\times10^{11}$M$_\odot$ for M31 from a
compilation by \cite{Mutch11}; $5.9\pm1.3\times10^{10}$M$_\odot$ for
the MW from \cite{Licquia16a}; and $4.5\pm1.5\times10^9$M$_\odot$ for
M33 from \cite{Corbelli03}.  The next three subsections details how
$\sigma_{\rm AD}$, $r_{50}$, $b_r$ and $\sigma_{r,0}$ are derived
for each galaxy.

\subsubsection{M31}
\label{sec:M31}

We use the measurements of stellar age and kinematics from two
comprehensive studies of M31 by \cite{Dorman15} and
\cite{Quirk19}. From \cite{Dorman15} we estimate $\sigma_{\rm LOS}$
from their Figures 16 at radii of 10 and 15 kpc, and also from Figure
12, where we adopt a characteristic radius of 12.5 kpc.
Parenthetically we note that inside of 8 kpc $\sim 1.25 r_{50}$ the
age-velocity dispersion abruptly changes, referring to
\cite{Dorman15}'s Figure 16. While there may remain a clear
distinction in the kinematics of the youngest and older populations,
the intermediate and older populations are indistinguishable in terms
of dispersion alone.

To convert the line-of-sight velocity dispersions ($\sigma_{\rm LOS}$)
to $\sigma_{\rm AD}$ we first scale $\sigma_{\rm LOS}$ to $\sigma_r$.
When the rotation curve is nearly flat and the epicycle approximation
(equation~\ref{eq:epi}) $\sim 1/2$, we can write $$\sigma_{\rm LOS} =
\sigma_r^2 [ (\sin^2(\phi) + \cos^2(\phi)/2)\sin^2(i) + \alpha^2
  \cos^2(i)].$$ Looking at the field coverage on M31 in
\cite{Dorman15}'s analysis, a reasonable estimate the characteristic
on-sky azimuthal angle sampled in this work is $\theta=45^\circ$,
which corresponds to $\phi\sim77^\circ$ for a disk inclination of
$i=77^\circ$. For $\alpha=0.6$ we find $\sigma_r = 1.03 \ \sigma_{\rm
  LOS}$.

To convert from $\sigma_r$ to $\sigma_{\rm AD}$ using
equation~\ref{eq:ad4} requires knowing the radial range of their
measurements with respect to M33's half-light radius.  For this
estimate we use $I$-band measurements and light profile decomposition
from \cite{Courteau11}, yielding bulge Sersic index of $n_S=2.2\pm0.3$ and
effective radius $R_e=1.0\pm0.2$ kpc, a disk scale-length of
$5.3\pm0.5$ kpc, and a disk-to-total luminosity ratio of
$0.75\pm0.04$. Numerically integrating the light profile we find
$r_{50}{\rm (M31)} = 6.5^{+1}_{-3}$ kpc. For an exponential disk we then
have $1<\sigma_{\rm AD}/\sigma_r < 2.6$ in the radial range of
interest between $1.5<r/r_{50}<2.3$.

\begin{figure}
 \centering
 \includegraphics[width=\columnwidth]{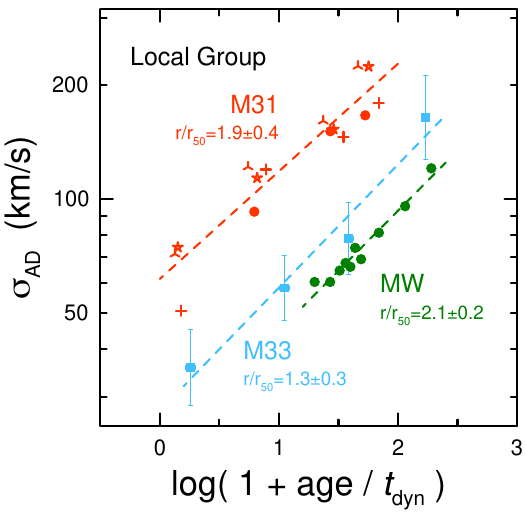}
   \caption{Trends of $\sigma_{\rm AD}$ versus the ratio of the stellar age to the local dynamical time for Local Group galaxies M31, the Milky Way and M33, constructed as described in the text. The radial range scaled to the half-light radius is indicated for each galaxy. Unweighted regressions for each galaxy are shown with dashed lines. For M31, circles represent asymmetric drift measurements from \citet{Quirk19}; three and four sided points are scaled from $\sigma_{\rm LOS}$ at 10 and 15 kpc from \citet{Dorman15}; five-pointed stars are from \citet{}{Dorman15}'s Figure 12. Data used for M33 and the MW are described in the text.}
   \label{fig:lg_ad}
 \end{figure}

\cite{Quirk19} directly measure asymmetric drift ($V_a$) which we
convert to $\sigma_{\rm AD}$ adopting $V_c = 250$ \kms. We adopt their
$V_a$ values derived using HI-envelope estimates of the circular
speed. It is interesting to note that $V_a<0$ for their youngest
stellar bin (main sequence stars with ages of 30 Myr). They attribute
this to star-forming regions originating from dense molecular clouds
that are dynamically colder than the ambient atomic and molecular gas.
Interestingly, \cite{Shetty20} also see evidence for such behavior in
their multi-age analysis of asymmetric drift of MaNGA galaxies, albeit
for a slightly older age stellar population and with respect to the
ionized gas tangential speed. However, the corresponding dispersions
for the same age population from \cite{Dorman15} seem, on the face of
it, to be inconsistent with this hypothesis for M31.  It may be that
velocity shear and molecular clouds lead to very rapid initial heating
of the in-plane dispersion, as suggested by \cite{Kokubo92}.
Resolving this tension no doubt will yield a better understanding of
heating processes in disks, particularly at early ages. For the
purposes of establishing the age--velocity dispersion relation for
M31, we exclude the \cite{Quirk19} $V_a$ value for the youngest age in
our analysis. The resulting data are plotted in
Figure~\ref{fig:lg_ad}, where there is good agreement between the
\cite{Dorman15} and \cite{Quirk19} values.  A simple linear regression
yields $b_r = 0.28\pm0.03$ and $\sigma_{r,0}=26\pm3$ \kms.

\subsubsection{Milky Way}
\label{sec:MW}

We use the measurements of stellar age and velocity dispersion for the
solar neighborhood from the Geneva-Copenhagen survey
\citep{Nordstrom04,Holmberg07,Holmberg09}. Specifically, we adopt the
data as presented in Figure 31 of \cite{Nordstrom04} for the radial
component dispersions, noting \cite{Seabroke07}'s caution against
interpreting dispersion values for in-plane components. 

To scale $\sigma_r$ to $\sigma_{\rm AD}$ we estimate $r_{50}$ by adopting estimates from \cite{Licquia16a} of a disk-to-total luminosity ratio of $0.84\pm0.03$ and a disk scale-length of $2.7\pm0.2$ kpc. We adopt a bulge Sersic index of 1.7 and a half-light radius of 1 kpc.  Numerically integrating the light profile we find $r_{50}{\rm (MW)} = 3.85 \pm 0.5$ kpc. This yields $\sigma_{\rm AD}/\sigma_r \sim 2.5$ at the solar circle, which we take to be at a radius of 8 kpc.  The resulting data are plotted in Figure~\ref{fig:lg_ad}. We neglect the errors associated with these measurements since they are small and relatively uniform. A simple linear regression yields $b_r = 0.32\pm0.03$ (essentially identical to values found by \cite{Nordstrom04}) and $\sigma_{r,0}=9\pm1$ \kms.

We also use more recent measurements of $b_r$ by \cite{Mackereth19} that cover a range of radii based on APOGEE-2 DR 14 and Gaia DR2 data (see references therein).

\subsubsection{M33}
\label{sec:M33}

We use measurements from \cite{Beasley15} for star-cluster ages and
velocity dispersions and adopt their inclination of $i=56^\circ$ for
M33's disk.  To convert the line-of-sight velocity dispersions
($\sigma_{\rm LOS}$) to $\sigma_{\rm AD}$ we first scale $\sigma_{\rm
  LOS}$ to $\sigma_r$.  Adopting the same approach as for M31, we find
the characteristic in-plane projection angle $\phi \sim 61^\circ$ for
M33, so that for $\alpha = 0.6$ we have $\sigma_r = 1.12 \sigma_{\rm
  LOS}$. To convert from $\sigma_r$ to $\sigma_{\rm AD}$ using
equation~\ref{eq:ad4} requires knowing the radial range of their
measurements with respect to M33's half-light radius.

\cite{Simon06} estimate the disk scale-length as $1.36\pm0.06$ kpc in
the $K$-band (adopting a distance of 800 kpc), which agrees well with
the earlier $K$-band measurements of \cite{Regan94}. \cite{Regan94}
tabulate scale-lengths in a wide range of pass-bands, indicating
significant color-gradients (the disk appears larger at bluer
wavelengths). However, there are some discrepancies in the tabulated
$I$-band value from \cite{Guidoni81} and the original work.  To match
to the $I$-band estimates for M31, we adopt \cite{Regan94}'s $J$-band
measurements with an uncertainty bracketed by their $K$-band value and
the original $I$-band value from \cite{Guidoni81}, corrected to the
same distance as \cite{Simon06}.  Assuming a pure exponential disk we
arrive at $r_{50}{\rm (M33)} = 2.4^{+0.5}_{-0.1}$ kpc. This places the
$\sigma_{\rm LOS}$ measurements from \cite{Beasley15} to fall between
$1<r/r_{50}<1.65$, yielding the corresponding scaling of
$1.6<\sigma_{\rm AD}/\sigma_r<2.2$. The resulting data are plotted in
Figure~\ref{fig:lg_ad}. A simple linear regression yields $b_r =
0.32\pm0.06$ and $\sigma_{r,0}=15\pm4$ \kms.

More recent work by \cite{Quirk22} based on photometric identification and spectroscopic measurements of field star kinematics in M33 shows little indication for the dramatic increase in asymmetric drift or velocity dispersion with stellar population age. I.e., their data is consistent with a nearly flat AVR.
The origin of the differences between these two studies is unclear, but it is relevant to note the related work by \cite{Gilbert22} who find M33 has a bimodal velocity distribution in its red-giant branch stellar population. \cite{Gilbert22} identify the hotter component as a compact halo population; when it is included in \cite{Quirk22}'s analysis, the AVR is no longer flat and more closely resembles the result from \cite{Beasley15}. Certainly in our MaNGA data no such distinction in populations is made. However, we do see evidence for some MaNGA galaxies of similar mass as M33 having flat AVR, and indeed one might argue the distribution in $b_r$ is bimodal below $\log {\rm M}_\star / {\rm M}_\odot h^{-2}_{70} = 10.4$. For present comparative purposes with MaNGA we continue with the results based on star-cluster measurements from \cite{Beasley15}.

\subsection{Comparison between MaNGA and the Local Group}
\label{sec:LG_comp}

\begin{figure}
   \includegraphics[width=\columnwidth]{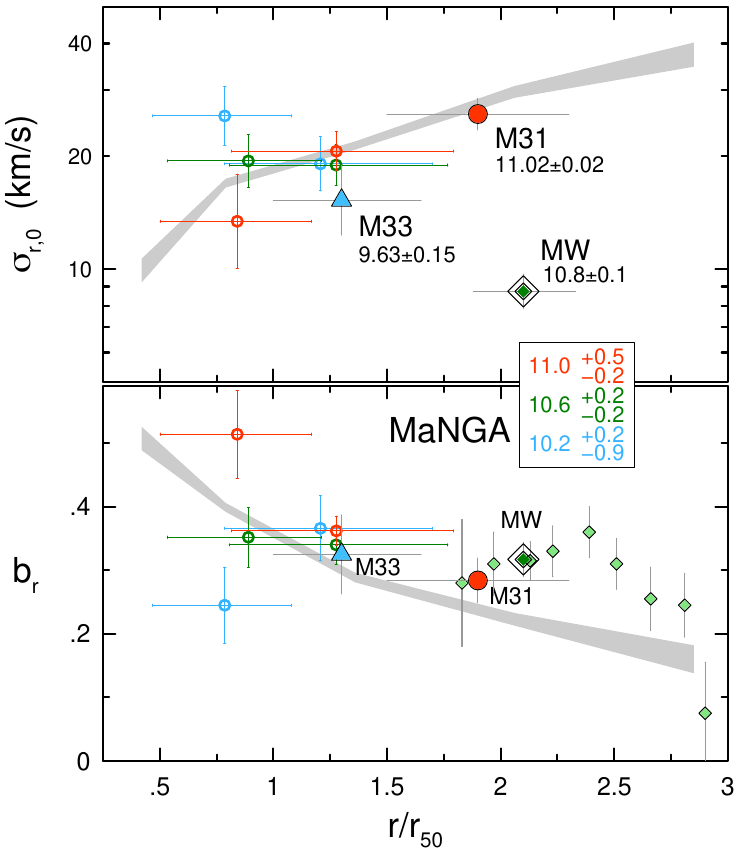}
   \caption{Asymmetric drift stratification model parameters $\sigma_{r,0}$ and $b_r$ versus radius scaled to $r_{50}$. MaNGA data (polygons and open dots) are shown for the  best-fitting model with both parameters free. Grey shaded polygons represent the ensemble method for all mass. Open red, green and blue points represent the indy method for the three mass bins defined in the key. Our estimates for values for M31, MW and M33 from the literature, as described in the text, are marked as filled red circles, green diamonds and light-blue triangles, respectively. Lighter green diamonds represent values for the MW from \citet{Mackereth19}.}
   \label{fig:tdyn_adm_comp}
\end{figure}

Figure~\ref{fig:tdyn_adm_comp} shows the stratification parameters for the MaNGA sample as a function of radius (i) for the indy-method results divided into three mass bins; and (ii) for the ensemble method for all masses combined. Keeping in mind that MaNGA measurements are based on asymmetric drift, these are compared to values derived for the three massive galaxies in the Local Group. 

There are clearly different radial ranges being probed in the two samples, with larger radii (with respect to the half-light radius) for the Local Group. Nonetheless, there is reasonable overlap near $1.5~r/r_{0.5}$ for the indy method, while the ensemble method extends over the full range sampled in the Local Group. The MW appears to be an outlier in having unusually low $\sigma_{r,0}$. While $\sigma_{r,0}$ is a particularly large extrapolation of the stratification model based on the data for the MW compared to M31 and M33, as seen in Figure~\ref{fig:lg_ad}, the same Figure does show that overall $\sigma_{\rm AD}$ is low for the MW at comparable values of stellar age/$t_{\rm dyn}$. M33 and M31 appear to follow the trend of increasing $\sigma_{r,0}$ with radius reasonably well. Both of these galaxies appear to have had significant interactions over the last 2-3 Gyr \citep{Putman09, DSouza18}. In light of this, and the fact that $\sigma_{r,0}$ is a model extrapolation for \textit{all} data considered here, $\sigma_{r,0}$ may not represent the true birth dispersion of stars. The characteristic stratification index values ($b_r$) for the Local Group are quite uniform, and while they do not show any clear mass trend, they are within the range of values and the radial trend seen in the MaNGA data.

%% file: conclusion.tex
\section{Summary and conclusion}
\label{sec:conclusion}

In this paper we have explored the age--velocity dispersion relation of galaxy disks in the MaNGA survey. Out of a sample of $\sim2800$ galaxies, we have selected a subset of $\sim500$ with regular kinematics indicative of rotating disks that have good tracers of tangential speed for both ionized gas and stars. We deemed these galaxies had suitably axisymmetric motions to characterize gas tangential speeds that could be related to the circular speed of the potential with modest corrections for pressure support. We have estimated and applied that correction to the gas kinematics from a relation between emission-line flux ratios and gas velocity dispersion calibrated at very high signal-to-noise \cite{Law21b}.

We calibrated a proxy for light-weighted mean ages, $t_{\rm LW}$, for unresolved stellar populations based on the \dnk~ spectral index. This estimator is both reliable and observationally robust, i.e., it can be measured directly in individual spectra and \textit{repeatably} in other data-sets. Galaxies in our sample contain stellar populations with $t_{\rm LW}$ between 0.2 and 10 Gyr.

We used a light-weighted measurement of the stellar tangential
speed to measure asymmetric drift with respect to the ionized gas. The
asymmetric drift signal scales with the amplitude of the stellar
velocity dispersion ellipsoid, but as a measure itself asymmetric
drift depends only on velocities. Using asymmetric drift as a proxy
for the stellar velocity dispersion alleviates the need to calibrate
dispersion measures that extend well below the instrumental resolution.
We showed that the asymmetric drift is proportional to in-plane
radial component $\sigma_r$ to within an estimatable factor of order
unity. We have also provided this proportionality. Values for $\sigma_{\rm
  AD}$ for our sample range from 20 to 300 \kms, which correspond to
$\sigma_r$ of roughly half these values.

We combined the sample age and dispersion measures to determine a mean age--velocity dispersion relation for the ensemble as well as for individual galaxies. Ensemble measurements were made at five characteristic radii with median values between 0.4 and 2.85~$r_{50}$, and in three bins of mass centered at $\log {\rm M}_\star / {\rm M}_\odot = 10.2$, 10.6 and 11.0. There is a modest bias between the light-weighting of the mean stellar population age and asymmetric drift signals. We have modeled this bias and corrected for it.

To characterize the AVR in the MaNGA sample, we have applied a simple power-law relation, $\sigma \propto t^b$, between velocity dispersion, $\sigma$, and the age of the stellar population, scaled by the local dynamical time measured for each galaxy at each radius. The two parameters of this model are the radial velocity dispersion of the stellar population and young ages, $\sigma_{r,0}$ and the power-law index (or stratification rate) $b_r$.  We find that barred galaxies in our sample yield erratic results for the derived stratification parameters, despite our visual selection for axisymmetric velocity fields. Consequently, we have removed barred galaxies (23\% of our sample) from tertiary analysis.

This analysis shows that stratification rates for \textit{un}barred galaxies tend to \textit{decrease} with radius and \textit{increases} with stellar mass, with values in the range $0.2<b_r<0.5$. These values of $b_r$ bracket the theoretical expectations for dynamical disk heating found in the literature. The values of $\sigma_{r,0}$ are anti-correlated, and are in the range of $10~\kms < \sigma_{0,r} < 25~\kms$.  Lower-mass galaxies ($\log {\rm M}_\star / {\rm M}_\odot  < 10.4$) appear to have little trend of their stratification parameters with radius. This transition is very close (in stellar mass) to  that noted by \cite{Dalcanton04} for galaxies to exhibit thin dust lanes. It is possible that changes in the mid-plane density of star-forming clouds at this mass transition play an important role in driving different trends in dynamical stratification; it is equally plausible that the inside-out mass-assembly and star-formation quenching processes evident in higher-mass galaxies are less pronounced at lower mass as a consequence.

A comparison of the stratification parameters of MaNGA galaxies to those derived from literature measurements for the three most massive galaxies in the Local Group (M33, the MW, and M31) show broadly consistent radial and mass trends. Hence dynamical stratification in MaNGA galaxies is broadly consistent with what we know about stratification in the Local Group. The stratification rate $b_r$ values for MaNGA galaxies are well within the range measured for the Milky Way, noting that for in-plane motions $b$ is smaller (and $\sigma_{0}$ is larger) compared to vertical motions \citep{Nordstrom04, Holmberg07, Holmberg09}. However, the MW appears to have a particularly small value of $\sigma_{r,0}$, and it is at the extrema of the range of values computed for galaxies of comparable mass in the MaNGA sample.

The correspondence of the best-fitting values in integrated star-light with those for resolved stellar populations in the Local Group indicates the method of using AD in integrated star-light is reliable. Nonetheless, we do find that the simple stratification model does not capture well the observed dispersion in the data, and we find modest systematics between fitting ensembles of galaxies compared to individual galaxies. This suggests that a richer analysis of high SNR data that is able to resolve the AVR in integrated star-light at specific locations \textit{within} galaxies \citep[e.g.,][]{Shetty20} will lead to progress in our understanding of the dynamical stratification of stellar populations.

%% file: appendix_apertures.tex
\section{Measurement apertures}
\label{app:app}

\begin{figure*}
 \centering
 \includegraphics[width=0.8\linewidth]{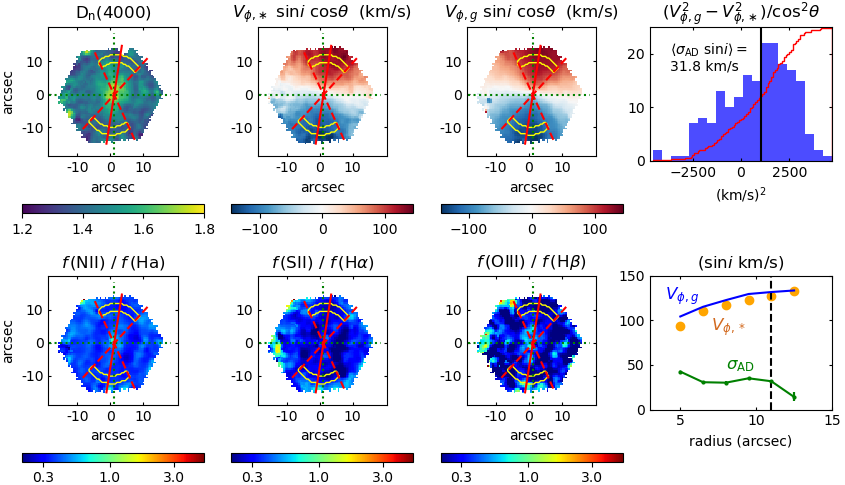}
  \caption{Example of kinematic and spectrophotometric signal extraction for 8146-12701 in bins having 2 arcsec radial width extending between $\pm40^\circ$ in $\theta$ about the major axis in the disk plane. In all map panels the kinematic major axis and $\pm40^\circ$ wedges are indicated with red lines; an example  aperture of 10-12 arcsec in radius is represented by a yellow contour. Top row (left to right): maps of \dnk, the full projected stellar and gas tangential speed ($V_{\phi,*}\sin i\cos\theta$ and $V_{\phi,g}\sin i\cos\theta$), and the histogram of $(V_{\phi,g}^2 - V_{g,*}^2)/\cos^2\theta$ -- the differences of the squared, azimuthally-deprojected gas and stellar tangential speeds for every pixel in the example bin. The red curve in the histogram panel shows the cumulative distribution; the black vertical lines gives the sample weighted mean. Bottom row (left to right): flux ratio maps of [NII] and [SII] to H$\alpha$, [OIII] to H$\beta$; and the radial profile of the inclination-projected kinematics ($V_{\phi,*}\sin i$, $V_{\phi,g}\sin i$, and $\sigma_{\rm AD}\sin i$) in the 6 radial bins with data for this galaxy. See text for further details.}
 \label{fig:app_example}
\end{figure*}

The extraction of kinematic and spectrophotometric measurements in apertures defined by radius and azimuth are illustrated in Figure~\ref{fig:app_example} for 8146-12701 (see also Figure~\ref{fig:examples_nobar}). The high-lighted radial aperture mid-point corresponds to 0.89$r_{50}$, or 6.4~kpc. To compute the asymmetric drift signal for this aperture we find the weighted mean of $(V_{\phi,g}^2 - V_{g,*}^2)/\cos^2\theta$ from the distribution measured for every spaxel, where $\cos^{-2}\theta$ corrects for the azimuthal projection of the tangential speed. Deprojecting for the measured inclination of $i=38\pm0.04$ deg yields $\sigma_{\rm AD}=51.6\pm2.9 (41.1)$~\kms, where the quantities in parentheses in this section are the standard deviation. Corrections for gas pressure (\S\ref{sec:gas_cor}) and light-weighted systematics (\S\ref{sec:lwa_sys}) yield $\sigma_{\rm AD}=53.8\pm3.9$~\kms\ and $\sigma_{\rm AD}=58.1\pm4.9$~\kms, respectively; the latter value is used in our subsequent analysis. Using the same weights and correcting for azimuth and inclination we find $V_{\phi,g}=213.8\pm0.4(6.1)$~\kms, which becomes $V_{\phi,g}=214.4\pm0.8$~\kms\ after correction for gas pressure. The trend in $V_{\phi,*}\sin i$, $V_{\phi,g}\sin i$, and $\sigma_{\rm AD}\sin i$ with radius for all apertures, before pressure and light-weighting corrections, are shown in the bottom right panel.

The weighted mean values for other quantities used in our analysis (\dnk\ and emission-line flux ratios) are found from the weighted mean of each quantity, also using the weights and mask from the computation of $(V_{\phi,g}^2 - V_{g,*}^2)/\cos^2\theta$. In the aperture centered at 0.89$r_{50}$ we have \dnk\ $= 1.389 \pm 0.003 (0.041)$,  f(NII)/f(H$\alpha) =0.346 \pm 0.002 (0.028)$, f(SII)/f(H$\alpha) = 0.273 \pm 0.003 (0.048)$ and f(OIII)/f(H$\beta) = 0.223 \pm 0.007 (0.093)$.

%% file: appendix_age.tex
\section{Correlation of \dnk~ with Light-Weighted Mean Age}
\label{app:age}

\begin{figure*}
 \centering
 \includegraphics[width=0.8\linewidth]{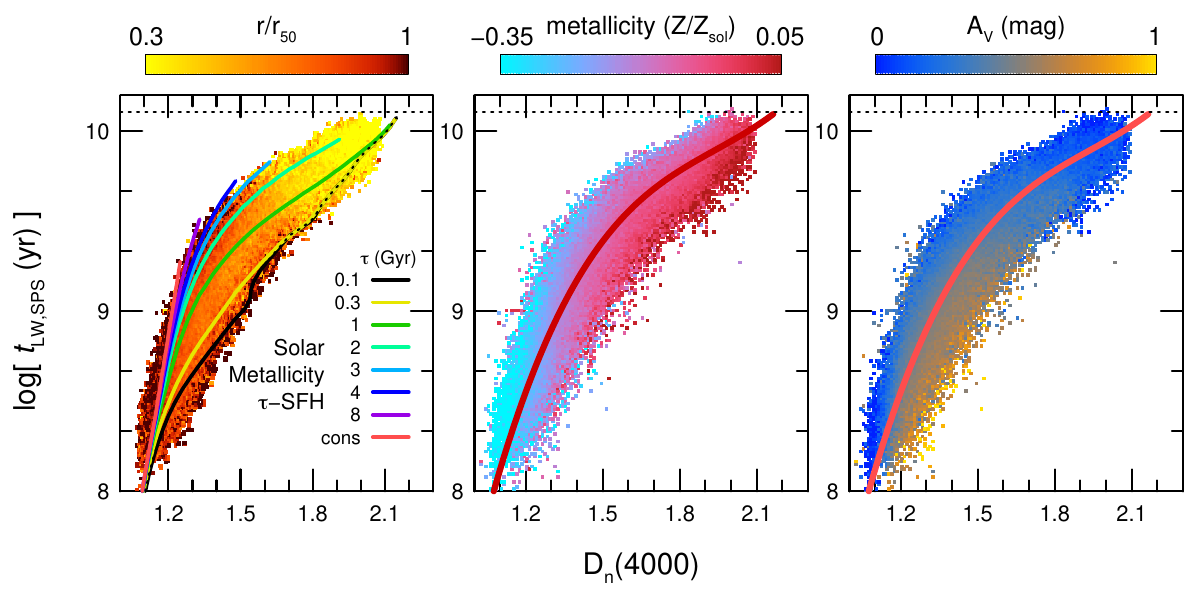}
  \caption{Light-weighted mean age ($t_{\rm LW}$, Johnson $V$ band)
    derived from stellar population synthesis (SPS) in Pipe3D versus
    \dnk~ as defined in the DAP and measured here for 184448 spectra
    in our sample.  Distributions are color-coded by median values of
    radius scaled to the half-light radius ($r_{50}$; left),
    metallicity (Z; middle), and extinction (A$_V$;
    right). Metallicity and exctinction are estimated from
    Pipe3D. Middle and right panels show our polynomial relation
    between age and \dnk. Comparisons are made in the left panel to
    composite populations \citep{BC03} with a range of
    exponentially-declining star-formation histories with e-folding
    time-scale $\tau$ at Solar metallicity.}
 \label{fig:age_dn4000_mod}
\end{figure*}

The correlation between \dnk~ and $V$-band light-weighted mean age
for exponentially declining star-formation histories (SFH) of solar
metallicity are shown in Figure~\ref{fig:age_dn4000_mod}. For these
models star-formation has progressed 12~Gyr.  While the SFH models
bracket the observed distribution, no single SFH well represents the
trend.  The bulk of the data are contained by SFH at solar metallicity
with $0.3<\tau({\rm Gyr})<3$, while $1<\tau({\rm Gyr})<2$ includes
most of the ridge-line. The dispersion about our adopted polynomial
relation between light-weighted age and \dnk~ is dominated by
variations in metallicity and extinction.

%% file: appendix_ad_svde.tex
\section{Derivation of relation between $\sigma_{\rm AD}$ and components of the stellar velocity ellipsoid}
\label{app:CBE}

We derive an equation for asymmetric drift that can be related to
observables in the MaNGA data set for Sersic density profiles and a
generalized rotation curve. With several more assumptions and
formalisms, we express this relation entirely in terms of factors
modulating a single component of the stellar velocity dispersion
ellipsoid (SVE). We choose here $\sigma_r$. We derive a specific case
where the density profile is exponential and the rotation curve is
well described by a radial tanh function. In what follows we define
the SVE vertical to radial axis ratio as $\alpha \equiv \sigma_z /
\sigma_r$.  While specific examples are common in the literature for
application to resolved stellar kinematics of the Milky Way's solar
neighborhood and several notable cases for application to external
galaxies \citep[e.g.,][]{Gerssen97,Gerssen00,Shapiro03,Noordermeer08,
  Weijmans08,Herrmann09a, Westfall11,Gerssen12,Westfall14} these
derivations often differ in their assumptions about which terms can be
ignored and are less general.

The quantity $\sigma_{\rm AD}$ (equation~\ref{eq:ad1}; asymmetric
drift) are related to components of the SVE and derivatives of
density and velocity distribution functions via integral moments of
the Collisionless Boltzman Equation (CBE). The rudiments of such a
relation are found in \cite{Binney08}. For tractable
application to observational data, simplifying assumptions are
required. As commonly done, we assume the galaxy stellar dynamics are
roughly in steady state, azimuthally symmetric, and have no radial
streaming motion. This allows us to eliminate terms such that the
integral of the radial-velocity moment ($V_r$) of the CBE yields \citep[e.g.][eq.~4.227]{Binney08}:
\begin{equation}\label{eq:app_ad2}
\sigma_{\rm AD}^2 =
\sigma_\phi^2 - \sigma_r^2 - \frac{r}{\rho}
\frac{\partial\rho\sigma_r^2}{\partial r} - r \frac{\partial
  \overline{V_r V_z}}{\partial z}.
\end{equation}

The first two terms can be consolidated in terms of, e.g.,
$\sigma_r$, by combining the integrals of the $V_r$ and $V_rV_\phi$
moments of the CBE to derive the epicyle approximation:
\begin{equation}\label{eq:epi}
\frac{\sigma_\phi^2}{\sigma_r^2} = \frac{1}{2} \left(\frac{\partial \ln
    V_\phi}{\partial \ln r} + 1\right).  
\end{equation}

The third term of equation~\ref{eq:app_ad2} can be parsed by assuming the
radial dependence of the tracer density ($\rho$) is proportional to
the ratio of the stellar surface-density ($\Sigma$) and the vertical
thickness ($h_z$): $\rho\propto\Sigma/h_z$. Dimensionally, we expect
$\rho\propto(\sigma/h)^2$, where $\sigma$ is a component of the SVE
and $h$ is a characteristic length-scale in that dimension. From this
one finds, e.g., $\rho\propto (\sigma_z/h_z)^2$ and the familiar
scaling $\sigma_z \propto \sqrt{\Sigma h_z}$. Indeed, for galaxies
with disks described by exponential light profiles
\cite{Martinsson13-VI} showed that $\sigma_z \propto \exp
(-r/2h_d)$. It would also be reasonable to expect $\sigma_r$ to have a
similar radial dependence, as seen in the Milky Way
\citep{Lewis89}. This happenstance is equivalent to $\alpha$ or the
relevant length-scales being constant with radius -- a further
assumption that we adopt. Generalizing for a Sersic surface-density
profile of index $n_S$ where $\Sigma(r) = \Sigma_{50} \exp( -b_n (
(r/r_{50})^{1/n_S}-1 ) )$ and $b_n \sim 2n_S-1/3$, the argument of the
derivative in the third term of equation~\ref{eq:app_ad2} can then be
written as a quantity proportional to $\exp
(-2b_n((r/r_{50})^{1/n_S}-1))$. The result can be expressed as
\begin{equation}\label{eq:cbe_term3}
-\frac{r}{\rho} \frac{\partial\rho\sigma_r^2}{\partial r} =
\sigma_r^2\left(\frac{2b_n}{n_S}\right)\left(\frac{r}{r_{50}}\right).
\end{equation}
This reduces to $\sigma_r^2(2r/h_d)$ for $n_S$ = 1, i.e., an
exponential disk.

\begin{figure}
  \centering
  \includegraphics[width=0.9\columnwidth]{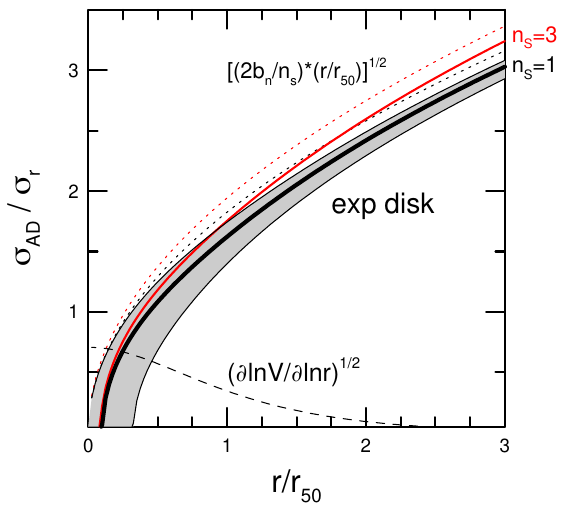}
  \caption{Ratio of $\sigma_{\rm AD}/\sigma_r$ versus radius scaled by
    the half-light radius. Values for an exponential disk ($n_S=1$)
    are given by black curves filled with grey that bound
    $0<\lambda<1$ and $0.3<\alpha<1$. The central, thick black curve
    adopts $\alpha=0.6$ and $\lambda=0.5$. Values for $n_S=3$, (also
    $\alpha=0.6$ and $\lambda=0.5$) are shown as a red solid
    curve. Values for the term in equation~\ref{eq:cbe_term3} for
    $n_S=1,3$ are shown as black and red dotted curves, respectively.
    Values for the term in equation~\ref{eq:drc} (with $h_{\rm
      rot}/r_{50}=0.6$) are shown as a dashed curves.}
  \label{fig:ad_scale}
\end{figure}

The last term represents the tilt in the SVE with respect to a
cylindrical coordinate system referenced to the disk. As shown by
\cite{Amendt91}, the derivative can be approximated as
\begin{equation}\label{eq:cbe_tilt}
\frac{\partial \overline{V_r V_z}}{\partial z} =
\lambda\left(\frac{\sigma_r^2-\sigma_z^2}{r}\right), 
\end{equation} 
where $0\leq\lambda\leq1$ and the extrema correspond to cylindrical
and spherical orientation, respectively.

These substitutions yield
\begin{multline}\label{eq:app_ad3}
\sigma_{\rm AD}^2 = \sigma_r^2 \biggl[
  \frac{1}{2}\frac{\partial\ln V_\phi}{\partial\ln r} + 
  \biggl(\frac{2b_n}{n_S}\biggr)\biggl(\frac{r}{r_{50}}\biggr) +
  \lambda(\alpha^2-1) - \frac{1}{2} \biggr],
\end{multline}
where the logarithmic derivative of the tangential speed can be
evaluated directly from the data or a parametric fit to the data. For
example, for a rotation curve well described by a model where $V(r) =
V_{\rm rot} \tanh (r/h_{\rm rot})$ we find 
\begin{equation}\label{eq:drc}
\frac{\partial\ln
  V_\phi}{\partial\ln r} = x \sech (x) \csch (x)
\end{equation}
where $x = r/h_{\rm rot}$. We adopt this form in our analysis here, which is suitable at all radii in a galaxy where the radial profile of disk mass and tangential velocity are well characterized by a Sersic profile of constant thickness and $\tanh$, respectively. The dominant term in equation~\ref{eq:app_ad3} is given by equation~\ref{eq:cbe_term3} for radii above $\sim 0.25 \ r_{50}$ for the full range of Sersic index and rotation curves in our sample. However this term is only weakly dependent on the Sersic index. The different terms in equation~\ref{eq:app_ad3} are illustrated in Figure~\ref{fig:ad_scale}.

%% file: appendix_lwa_systematics.tex
\section{Systematics between light-weighted ages and kinematics}
\label{app:lwa_sys}

\begin{figure*}
  \centering \includegraphics[width=0.95\textwidth]{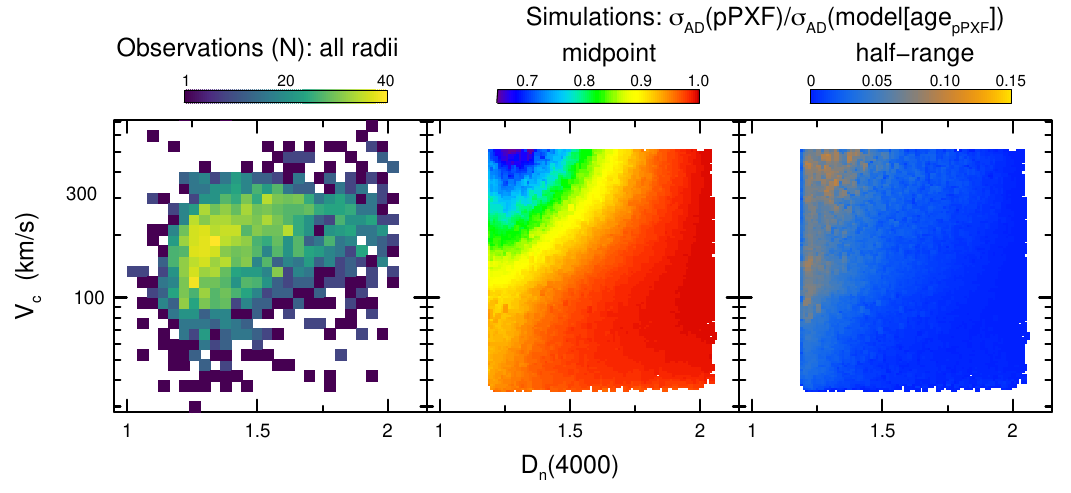} 
  \caption{Bivariate distribution as a function of rotation speed
  ($V_c$) and \dnk\ of observations (number per bin, left panel) and
  systematics in measured $\sigma_{\rm AD}$ due to differential
  light-weighting of kinematics and ages from full-spectrum fitting
  (middle and right) as a function of rotation speed ($V_c$) and \dnk.
  Systematics are derived from simulations described in the text. The
  midpoint (middle panel) and half-range over the midpoint (right panel) provide
  robust estimates of the mean and width of the simulations
  distribution.}
\label{fig:lwa_sys}
\end{figure*}

To quantify potential mismatch between light-weighted ages and kinematics from full-spectrum fitting over the same range in wavelength, we created $5\times10^4$ mock spectra spanning a wide range of star-formation histories (SFH; $\psi(t)$) and rates of dynamical stratification that appear to span what is observed in our data.

The mock spectra are composites of solar metallicity
SSPs \citep[MIUSCAT,][]{Vazdekisetal2012}, rendered at the MaNGA
spectral resolution, weighted according to the SFH, and {\it each}
velocity broadened and shifted according to their age and the
dynamical stratification model. We adopt the power-law dynamical
stratification for $\sigma_r$ exactly of the form of
equation~\ref{eq:sor} with $0.05<b_r<0.5$ and $5<\sigma_{r,0}
\ (\kms)<50$.  Model dynamical times ranged from 0.013 to 0.13 Gyr
with a median of 0.042 Gyr, closely matching the observed median
dynamical time of 0.035 Gyr, ranging from 0.02 Gyr at $0.25~r_{50}$ to
0.05 Gyr at $1.25~r_{50}$ (90\% of the sample has dynamical times
between 0.011 and 0.077 Gyr over this radial range). The velocity
scale of each simulation is set by $V_c^2 = 35^2 + \sigma_{\rm
AD}^2(t=14{\rm Gyr})$ where $\sigma_{\rm AD} = 1.62~\sigma_r$ (i.e.,
$r=r_{50}$, although this choice is not important).  The $V_c$
zeropoint value of 35 \kms\ is introduced to ensure the oldest stellar
populations retain some rotational support as part of the galaxy disk
population. SSPs are broadened by $\sigma_r(t)$ and shifted in
velocity by $(V_c^2-\sigma_{\rm AD}(t)^2)^{1/2}$, following
equation~\ref{eq:ad1}.

Following studies of disk galaxies, e.g., the Milky Way
\citep{Pilyugin96a}, we take the star-formation rate $\psi(t)$ to be smooth and of the form:
\begin{equation}
\label{eq:sfh}
\begin{split}
\psi(t) & \propto  t \exp(-t/t_{\rm top}), \ t<t_{\rm top} \\
 & \propto \exp(-t/\tau_{SF}), \ t>t_{\rm top},
\end{split}
\end{equation}
with $t$ the age since the first stars formed. While actual star-forming histories are not always smooth and in cases include bursts, modeling such stochasticity is beyond the current scope. The above $\psi(t)$ consists of a rising rate up to $t_{\rm top}$, and either a flat or exponentially declining rate thereafter. We match the observed range in \dnk\ and age by letting $0.5<t_{\rm top}{\rm (Gyr)}<8$, $-0.5<\tau_{SF}^{-1} {\rm (Gyr^{-1})}<1$, and allowing formation epochs to span from 2.1 and 10.05 Gyr after the Big Bang. This corresponds to formation redshifts of $0.3<z<3$. While this range extends to recent times, keep in mind that these simulations are meant to mimic specific locations within galaxy disks, the outskirts of which may begin to form relatively late. Taking the age of the universe to be 14 Gyr, some simulation are still in the rising portion of their SFH today. The simulation values considered are for the current epoch ($z=0$), suitable for comparison to the MaNGA observations of galaxies with modest look-back times under 1 Gyr.

To match simulation analysis to our data analysis we use full spectrum
fitting with pPXF, with no regularization, to recover the velocities
and light-weighted ages. The same SSPs that form the mocks are adopted
as templates. This eliminates the effect that template mismatch might
have on the measured kinematics and ages; in the case of velocities we
find template mismatch is a very modest effect \citep[$<6$ \kms;][]{Shetty20}.

We quantify the systematic in terms of the ratio of the measured $\sigma_{\rm AD}$ from pPXF ($\sigma_{\rm AD}{\rm (pPXF)}$) to the quantity estimated from the model $\sigma_r(t)$ at the light-weighted age derived from the pPXF fitting ($\sigma_{\rm AD}{\rm (model[age_{pPXF}])}$). In this way we estimate the correction for $\sigma_{\rm AD}$ appropriate for the pPXF-measured age.

The amplitude of the $\sigma_{\rm AD}$ systematic is shown in
Figure~\ref{fig:lwa_sys} as a bivariate function of $V_c$ and \dnk (middle),
and compared to the distribution of data in our sample at all radii (left).
The middle map serves as the look-up table to correct
$\sigma_{\rm AD}$ as a function of the observed \dnk\ and $V_c$
derived from equation 3.